\documentclass[useAMS,fleqn,usenatbib]{mnras}

\usepackage{newtxtext,newtxmath}

\usepackage[T1]{fontenc}
\usepackage{ae,aecompl}

\usepackage{graphicx}	
\usepackage{amsmath}	
\usepackage{amssymb}	

\usepackage{subfig}
\usepackage{epsfig}
\usepackage{times}
\usepackage{natbib}
\usepackage{amsfonts}

\newif\ifAMStwofonts

\newcommand\rin{$r_{\rm m}$}

\newcommand\rc{$r_{\rm c}$}

\newcommand\teq{$T_{\rm eq}$}
\newcommand\peq{$P_{\rm eq}$}
\newcommand\teqo{$T_{\rm eq,0}$}
\newcommand\peqo{$P_{\rm eq,0}$}
\newcommand\Mout{$\langle\dot{M}\rangle_{\rm out}$}
\newcommand\Mavg{$\langle\dot{M}\rangle$}

\title[Spin equilibrium in strongly-magnetized accreting stars]{Spin equilibrium in strongly-magnetized accreting stars}

\author[C. R. D'Angelo]{
C. R. D'Angelo$^{1}$\thanks{E-mail: dangelo@strw.leidenuniv.nl}
\\
$^1$Leiden Observatory, Leiden University, Postbus 9513, 2300 RA
 Leiden, The Netherlands
}

\date{Accepted XXX. Received YYY; in original form ZZZ}

\pubyear{2016}

\begin{document}
\label{firstpage}
\pagerange{\pageref{firstpage}--\pageref{lastpage}}
\maketitle

\begin{abstract}
  Strongly magnetized accreting stars are often hypothesized to be in
  `spin equilibrium' with their surrounding accretion flows, which
  requires that the accretion rate changes more slowly than it takes
  the star to reach spin equilibrium. This is not true for most
  magnetically accreting stars, which have strongly variable accretion
  outbursts on time-scales much shorter than the time it would take to
  reach spin equilibrium. This paper examines how accretion outbursts
  affect the time a star takes to reach spin equilibrium and its final
  equilibrium spin period. I consider several different models for
  angular momentum loss -- either carried away in an outflow, lost to
  a stellar wind, or transferred back to the accretion disc (the
  `trapped disc').  For transient sources, the outflow scenario leads
  to significantly longer times to reach spin equilibrium ($\sim$10x),
  and shorter equilibrium spin periods than would be expected from
  spin equilibrium arguments, while the `trapped disc' does not. The
  results suggest that disc trapping plays a significant role in the
  spin evolution of strongly magnetic stars, with some caveats for
  young stellar objects.
\end{abstract}




\section{Introduction}

The spin rate of a star is strongly affected by the presence of a
magnetic field. Neutron stars, for example, show a clear inverse
correlation between magnetic field strength and spin rate: the fastest
millisecond pulsars ($P_{\rm spin} \sim 0.002$--$0.005~{\rm s}$) have
a typical field strength of $B\sim 10^8~{\rm G}$, while magnetars with
$B\sim 10^{14}~{\rm G}$ have typical spin periods of a few
seconds. The influence of the magnetic field is even stronger when
such stars are accreting gas. Although accreted gas adds considerable
angular momentum to the star, accreting magnetized stars generally
spin well below their break-up velocity (sometimes many orders of
magnitude slower), indicating that the presence of the magnetic field
is able to regulate the transport of angular momentum between the star
and surrounding gas.

The stellar magnetic field in fact strongly affects the dynamics of
the accreting gas, and couples the star to its surrounding
environment. Close to the star, matter is forced to flow along field
lines on to the magnetic poles. At the boundary of this region
(typically called the {\it magnetospheric} or {\it Alfv\'en radius},
\rin), the magnetic field can in turn be significantly distorted by
the gas, which exerts a torque on the star.  The sign of the torque
depends on the relative location between \rin\ and the {\it
  co-rotation radius}, $r_{\rm c} \equiv (GM/\Omega^2_*)^{1/3}$, or
the location where a Keplerian disc co-rotates with the star. If
$r_{\rm m} > r_{\rm c}$, the star spins faster than the inner disc, so
that field lines coupling the two will gradually spin down the
star. The location of the magnetospheric radius itself is chiefly
determined by the stellar magnetic field and accretion rate, although
it is also sensitive to the detailed interaction between the gas and
the magnetic field.

This basic picture leads naturally to the concept of `spin
equilibrium' (or `disc locking' in young stars), whereby the star's spin
rate gradually adjusts itself until the net torque on the star is
roughly zero and the accretion flow is truncated near the co-rotation
radius, $r_{\rm m} \simeq r_{\rm c}$.  In this way the star's dipolar
magnetic field can be estimated, provided the spin and accretion rate
are known. Assuming spin equilibrium is reached requires assuming a
steady mass accretion rate -- i.e. that the timescale on which the
accretion rate changes is generally much longer than the `spin
equilibrium time' (\teq), defined as the time the star takes to reach
its `spin equilibrium period' (\peq).

It is not clear that this assumption is widely valid for accreting
magnetized stars, either compact stars (magnetized white dwarfs and
neutron stars) or young stellar objects (YSOs). Most magnetized
compact stars in binary systems are transient, showing short accretion
outbursts followed by long periods of quiescence. In weak-field
accreting neutron stars, the low-mass X-ray binaries (LMXBs), the
observed duty cycle is on average 3 per cent, but can be well below 1
per cent when allowing for the limited observing baseline
\citep{2015ApJ...805...87Y}. The luminosity difference in LMXBs
between outburst and quiescence can span many orders of magnitude,
suggesting a huge change in accretion rate. High magnetic field
transient neutron stars can also show strong variability. In one
particular class of system, Be X-ray binaries (neutron stars that
accrete from the wind or disc surrounding a Be star), the duty cycles
are $\sim 5-20$ per cent
\citep{2011Ap&SS.332....1R,2014MNRAS.437.3863K}, and the dynamic range
can be 1--5 orders of magnitude between outburst and quiescence.

At least some young stellar objects (YSOs) also show large-scale
variability, but its prevalence is much harder to constrain, since
they evolve on much longer time-scales than compact binaries. The most
dramatic accretion outbursts are FU Ori-type outbursts, where the
luminosity increases by $\sim 1000$ \citep{1996ARA&A..34..207H},
with an outburst duration of at least decades and recurrence time of
several thousand years. Strong luminosity variations of about 1--2
orders of magnitude on shorter (years) time-scales are also sometimes
seen (the `EXor' class; \citealt{2014prpl.conf..387A}, suggesting that
the mean accretion rate can vary considerably at different points in
the TTauri phase.

Variations in accretion rate may help explain observations in
magnetospherically accreting systems that do not easily fit in to the
standard spin equilibrium picture. YSOs with discs, for example, show
clear indications of magnetic field regulated rotation, spinning well
below their break-up values. However, attempts to confirm disc locking
have been mixed, or seemed to contradict simple model predictions
\citep{2012ApJ...756...68C}. Another example: a recent survey of the
spin-rates in Be X-ray binaries found that the neutron star frequently
rotates much more slowly than would be expected for a moderate
($10^{12}$~G) magnetic field star in spin equilibrium, suggesting that
much larger fields ($10^{14}$--$10^{15}$~G) are present
\citep{2014MNRAS.437.3863K}. The large number of such binaries makes
this unlikely from a population point of view, and it also seems to
contradict magnetic field estimates from cyclotron lines in analogous
Galactic systems with similar spin rates and luminosities \citep{2014MNRAS.437.3664H}.

This paper investigates how large-amplitude, short-timescale accretion
rate variations affect the spin evolution of the star, and how this
evolution changes for different models for stellar angular momentum
loss. As described in more detail below, it is not clear whether most
angular momentum is lost through stellar outflows (winds from the
star, or at the disc--magnetic field interface) or whether angular
momentum is mainly lost to the accretion disc. As I demonstrate below,
different angular momentum loss mechanisms lead to different
predictions for spin evolution as a function of accretion rate, so
that comparing the long-term spin evolution of each model with
different accretion rate profiles may offer new observational tests to
distinguish between them.

\section{Models for magnetospheric accretion}
\label{sec:mag_acc}

The basic picture for how a strong stellar magnetic field interacts
with accreting gas is theoretically fairly well established and
supported by numerical magnetohydrodynamical (MHD) simulations,
although there are still some significant uncertainties (see
e.g. \citealt{2004Ap&SS.292..573U} or \citealt{2014EPJWC..6401001L}
for theoretical reviews).

In the region closest to the star, the magnetic field completely
determines the gas behaviour, truncating the accretion disc at the
magnetospheric radius, \rin. The field lines in this inner region
remain closed and infalling gas flows along the field lines to accrete
near the magnetic poles of the star. Just outside \rin, field lines
couple to the disc, and this coupling exerts a torque on the star due
to the differential rotation between the disc and the star, which
twists magnetic field lines and generates a toroidal field
component. In the low-density atmosphere above the disc, `force-free'
conditions apply, meaning that the increasing magnetic pressure (from
the generated toroidal field component) causes magnetic field lines to
inflate and eventually open up, potentially driving an outflow from
the disc (e.g. \citealt{1997ApJ...489..199G,
  1997ApJ...489..890M}). Some open field lines may subsequently
reconnect, and small-scale instabilities at the interface between the
disc and the closed magnetosphere can recouple the star and the disc,
starting the cycle again.

The resulting global field geometry is significantly different from
early suggestions
(e.g. \citealt{1977ApJ...217..578G,1979ApJ...232..259G}; hereafter GL)
in that only a small region at the inner edge of the disc is coupled
to the magnetic field ($\Delta r/r < 1$). In contrast the GL model
proposed that stellar magnetic field lines remain embedded over a wide
radial extent in the disc, so that large amounts of angular momentum
are transported to the disc through the twisting of the field
lines. This was shown to be physically inconsistent by
\cite{1987A&A...183..257W}, since the high level of twist proposed by
this model would be enough to completely disrupt the outer disc. Later
work (e.g. \citealt{1990A&A...227..473A,
  1995MNRAS.275..244L,1996ApJ...468L..37H}) demonstrated that field
lines will tend to become open, so that only the inner edge of the
disc is coupled to the star.

 All accreting magnetic stars are generically observed to spin well
 below their break-up frequencies, some (e.g. some neutron stars with
 Be star or giant companions) up to six orders of magnitude more
 slowly. As the star accretes from the truncated disc, the angular
 momentum in the gas will be added to the star and make it spin
 faster, but how the star sheds angular momentum remains
 uncertain. MHD simulations tend to show strongly time-dependent
 accretion and outflows that carry away angular momentum, although the
 details remain simulation dependent (compare
 e.g. \citealt{2013A&A...550A..99Z,2014MNRAS.441...86L}). Simulations
 can also show strongly distorted field lines around the rotation
 axis, which carry away a significant amount of angular momentum from
 the star. Angular momentum can also be deposited directly into the
 accretion disc, changing its structure (see
 Section~\ref{sec:trapped};
 \citealt{1977PAZh....3..262S,2010MNRAS.406.1208D}), or be removed via
 a wind from the stellar surface (Section~\ref{sec:wind};
 \citealt{2005ApJ...632L.135M}).

\subsection{Location of magnetospheric radius}
\label{sec:rm}
The location of the disc's inner edge can be estimated from the
accretion rate and the star's magnetic field and mass. For the simple
case in which gas accretes radially on to the star, \rin\ is estimated
by setting the ram pressure of the infalling gas $\rho \varv^2$ equal
to the magnetic pressure of the dipolar magnetic field $B^2/8\pi$. In
terms of the mass accretion rate, this leads to a `standard'
expression for \rin\ \citep{1972A&A....21....1P}:
\begin{equation}
  \label{eq:rmGL}
  r_{\rm m,0} = \mu^{4/7}\dot{M}^{-2/7}(2GM_*)^{-1/7},
\end{equation}
where $\dot{M}$ is the accretion rate through the inner regions of the
disc, $\mu = B_*R _*^3$ is the magnetic moment of the star, and $M_*$
is its mass. For accretion from a circumstellar disc, the thinness of
the disc and the Keplerian rotation makes it more difficult for the
magnetic field to force the gas into corotation with the star, so that
\rin\ is smaller than for the radial-infall case
(e.g. \citealt{1979ApJ...232..259G,1987A&A...183..257W}); $r_{\rm m} =
\xi r_{\rm m,0}$; $\xi< 1$. In this case \rin\ is very sensitive to
the details of the coupling between the accretion disc and magnetic
field, which is the most uncertain aspect of the problem. Various
revised theoretical estimates for \rin\ have been proposed, suggesting
$\xi \sim 0.5-1$
(e.g. \citealt{1977ApJ...217..578G,1993ApJ...402..593S}; hereafter
ST93,\citealt{1996ApJ...465L.111W,2007ApJ...671.1990K,2008A&A...478..155B}.

\cite{1987A&A...183..257W} suggested a slightly different approach in
estimating \rin\ in a disc by imposing conservation of angular
momentum flux across \rin. This is most significant when the rotation
rate of the inner disc is close to the star's rotation rate. Using
this estimate gives a slightly different expression from
equation~(\ref{eq:rmGL}), and explicitly incorporates the star's
rotation frequency, $\Omega_*$:
\begin{equation}
  \label{eq:rmST}
r_{\rm m} = \left(\frac{\eta}{4}\right)^{1/5}\mu^{2/5}\Omega_*^{-1/5}\dot{M}^{-1/5}.
\end{equation}
Here $\eta = B_{\phi}/B_{\rm r} < 1$ is the magnitude of the toroidal
magnetic field, $B_{\phi}$ generated by twisting magnetic field lines
through differential rotation between the star and the disc. For
$r_{\rm m} = r_{\rm c}$, equation~(\ref{eq:rmST}) reduces to
equation~(\ref{eq:rmGL}) with $\xi \sim 0.4-0.7$ (for $\eta =
0.1-1$). Although the location of \rin\ is only uncertain by a factor
of a few, the strong dependence of $\dot{M}$ on \rin\ means that the
{\it accretion rate} for which $r_{\rm m} = r_{\rm c}$ (i.e. when spin
equilibrium is reached) can be uncertain by up to $\xi^{-7/2} =
300\times$.

The geometrical structure of the accretion disc can also affect the
location of \rin.  At very low or high accretion rates, the standard
`thin disc' accretion solution \citep{1973A&A....24..337S} likely does
not apply, and the flow becomes a geometrically thick
radiatively-inefficient accretion flow (RIAF, such as the ADAF
solution; \citealt{1994ApJ...428L..13N}). A RIAF rotates with
significantly sub-Keplerian velocities and is much less dense than a
thin disc at the same accretion rate. This means that for the same
accretion rate, the magnetospheric radius will likely be larger for a
RIAF than a thin disc.  This suggests that a significant change in
accretion flow structure and geometry may be accompanied by a large
change in \rin\ even without changing $\dot{M}$, which could have a
strong observational effect. Additionally, since the accreted gas is
considerably sub-Keplerian and geometrically thick, the angular
momentum exchange between the flow and the star could be considerably
altered, driving stronger outflows, for example, or enhancing the spin
down rate of the star. The interaction between a RIAF and a magnetic
field has not been studied in detail, but is likely very relevant both
for LMXBs at low luminosity and neutron stars accreting at
super-Eddington rates, such as the recently discovered pulsars in
ultra-luminous X-ray binaries \citep{2014Natur.514..202B,
  2016ApJ...831L..14F}.

\subsection{Standard accretion/ejection model}
\label{sec:AccEj}

The simplest model for angular momentum loss proposes that infalling
gas (and its angular momentum, $\sim \dot{M} \Omega_K(r_{\rm m})r_{\rm
  m}^2$) is either accreted on to the star (when $r_{\rm m} < r_{\rm
  c}$) or ejected in an outflow (when $r_{\rm m} > r_{\rm c}$).  If
all the specific angular momentum of the gas is either accreted or
expelled, the rate of angular momentum change in the star (i.e. the
angular momentum of the accreting gas, using equation~(\ref{eq:rmGL})
for the location of \rin) is given by:
\begin{equation}
  \label{eq:jdotAE}
  \dot{J} = 2^{-1/14}\xi^{1/2}\dot{M}^{6/7}\mu^{2/7}(GM_*)^{3/7}\tanh\left(\frac{r_{\rm m}-r_{\rm c}}{\Delta r_2}\right).
\end{equation}
Here the $\tanh$ function and the parameter $\Delta r_2$ are
introduced to move smoothly between the two solutions (ejection
versus accretion). Simulations typically show that the transition between
accretion and strong outflow occurs across a wide range of accretion
rates (which sets \rin), so that there are a range of accretion rates
that show both accretion and ejection
(e.g. \citealt{2003ApJ...588..400R}). By introducing $\Delta r_2$ I
can systematically investigate how the size of this transition
lengthscale affects the final equilibrium spin period and spin-down
time of the star (see also in Section \ref{sec:transition}).

Some simulations (e.g. \citealt{2013A&A...550A..99Z}) show very
energetic outflows, in which matter is ejected well above its escape
velocity, so that rate of angular momentum loss is larger than $\sim
\dot{M}_{\rm ej} (GMr_{\rm m})^{1/2}$ (where $\dot{M}_{\rm ej}$ is the
mass loss rate of the outflow).  This will increase the equilibrium
$\dot{M}$ (the accretion rate where $r_{\rm m} = r_{\rm c}$ since more
gas will reach the stellar surface without spinning up the star.
These simulations also show mass ejections even during phases
dominated by accretion, demonstrating that there can be significant
angular momentum lost from the star even during accretion phases.
The increase in the equilibrium $\dot{M}$ can be approximated by
changing $\xi$ in equation~(\ref{eq:rmGL}). For simplicity however, I
make the explicit assumption that in the limit $r_{\rm m} \gg r_{\rm
  c}$ (the strong propeller regime), the outflow of angular momentum
is limited by $\dot{M}_{\rm ej}(GMr_{\rm m})^{1/2}$.

In equation~(\ref{eq:jdotAE}) and throughout this paper, the accretion
rate $\dot{M}$ refers to the amount of gas accreting through the inner
regions of the disc. If most of this gas is ejected, the net accretion
rate on to the star will naturally be much lower. For a disc
magnetically truncated at more than a few stellar radii from the star,
the stellar accretion rate largely determines the accretion
luminosity, so it is somewhat difficult to define the accretion rate
through the disc without an accretion model: is their low luminosity
because most of the gas is begin expelled (the ejection scenario) or
because the accretion rate is intrinsically low but accretion
continues fairly efficiently (the trapped disc scenario outlined in
Section \ref{sec:trapped}).

\subsection{Spin regulation by a stellar wind}
\label{sec:wind}
It has also been suggested (specifically for young stars) that angular
momentum could be lost from a stellar wind powered by accretion energy
\citep{2005ApJ...632L.135M}. As described by
\cite{2005ApJ...632L.135M}, the angular momentum loss to the wind is
given by:
\begin{equation}
\label{eq:jdotWIND}
\dot{J} = -\dot{M}_w\Omega_*r^2_{A},
\end{equation}
where $\dot{M}_w$ is the mass loss rate in the wind, and $r_{\rm A}$
is defined as the location where the wind speed equals that of
magnetic Alfv\'en waves:
\begin{equation}
r_{\rm A} \sim R_* K \left(\frac{\mu^2}{\dot{M}_{w}\varv_{\rm esc} R_* }\right)^{m}.
\end{equation}
Here, $K \sim 2.1$ and $m\sim 0.2$ are fit constants from MHD simulations
and $\varv_{\rm esc}$ is the escape speed at the stellar surface
\citep{2008ApJ...681..391M}. The distinction in terms of angular
momentum loss between this and the accretion/ejection picture is that
here the star is assumed to efficiently lose angular momentum to the
wind at all accretion rates, instead of only when there is a
significant centrifugal barrier (i.e. $r_{\rm in} > r_{\rm c}$).

Whether a wind can efficiently carry away angular momentum thus
depends largely on the amount of mass loss in the wind (assuming it is
launched not far above its escape velocity). YSOs are observed to have
outflows of up to $\sim 10$ per cent of $\dot{M}$ for protostellar
systems \citep{2005ApJ...632L.135M}, but it is difficult to tell
whether this outflow originates from the star or the inner disc.
The ability of a stellar wind to efficiently carry away enough angular
momentum to regulate YSO spins has further been challenged by
\cite{2011ApJ...727L..22Z}. I assume in this paper that a wind
acts in conjunction with the accretion/ejection model, so that at low
$\dot{M}$ there are two sources of angular momentum loss: from the
wind and from a centrifugally launched outflow.

\subsection{Trapped disc model}
\label{sec:trapped}
When the inner edge of the accretion disc lies outside the corotation
radius, a centrifugal barrier inhibits accretion on to the
star. However, the disc--field interaction may not be strong enough to
drive a strong outflow.  If instead the disc--field interaction adds a
considerable amount of angular momentum to the inner disc, the disc
density structure and \rin\ will become `trapped' close to \rc, so
that even as $\dot{M}$ decreases the disc will stay near \rc\ and
accretion will continue
(\citealt{1977PAZh....3..262S,2010MNRAS.406.1208D}, hereafter DS10;
\citealt{2012MNRAS.420..416D}). As a result, the spin down rate as a
function of accretion rate will be significantly different from
equation~(\ref{eq:jdotAE}) or (\ref{eq:jdotWIND}). For a normal
accretion disc, this condition will at least be true for
$r_{\rm m} < 1.3 r_{\rm c}$, since the energy available through
differential rotation between the field and disc is not enough to
expel gas at a rate that matches the accretion rate in the
disc. Depending on how efficiently gas can be loaded into an outflow,
this situation can also apply for larger truncation radii.

However, once a trapped disc has formed, the inner edge of the disc is
no longer given by an equation of the form
equation~(\ref{eq:rmGL}). Instead, \rin\ is almost independent of
$\dot{M}$ and is determined by balancing the torque from the
disc--field interaction, $\tau_B$, with torque transmitted outwards by
viscous stress in the disc, so that:
\begin{equation}
3\pi\nu\Sigma(r_{\rm m})r^2\Omega_{\rm K}(r_{\rm m})=\tau_B = \eta r^2\Delta r B^2,
\label{eq:vt}
\end{equation}
where $\nu$ and $\Sigma$ are respectively the effective viscosity and
surface density of the accretion disc. In consequence, the accretion rate through
the inner disc can decrease to a very low rate or even drop to zero,
but the inner disc edge will never move very far from \rc.

DS10 and \cite{2011MNRAS.416..893D,2012MNRAS.420..416D} studied accretion
disc behaviour in these conditions, and found that accretion proceeded
either continuously or in short accretion bursts (much faster and
weaker than full accretion outbursts). Depending on the strength of
the coupling between the field and the disc, they also found that the
star could be efficiently spun down by the presence of a disc, even
when the accretion rate is extremely low.  \cite{2012MNRAS.420..416D}
called this state a `trapped disc', because the inner disc edge
remains trapped close to \rc\ as the average $\dot{M}$ through 
the outer disc decreases.

Spin regulation in the trapped disc model superficially resembles the
model suggested by GL, since accretion on to the star continues even
though the star is being spun down. However, it is fundamentally
different, in that it incorporates a more physically realistic,
potentially non-steady picture for the interaction between the disc
and the magnetic field coupling region, rather than the steady-state,
extended region of coupled field lines in GL. Rather than focus in
detail on how the disc and the magnetic field couple (which
simulations show is likely a complicated and non-steady process), DS10
instead assumed that the disc--field interaction adds angular momentum
to the disc, and demonstrated how the disc structure changes as a
result of this interaction. Furthermore, since the DS10 model has a
self-consistent description for the disc as a function of accretion
rate for all accretion rates, it does not break down at low $\dot{M}$
like the model of GL (which has no steady accretion solutions $r_{\rm
  m} > r_{\rm c}$).

The rotating magnetic field provides an additional spin-down torque
(comparable to angular momentum loss from a rotating magnetic dipole
in vacuum):
\begin{equation}
  \label{eq:pulsar}
  \dot{J} = -\frac{2\mu^2\Omega^3}{3c^2}.
\end{equation}
Except for weak-field accreting millisecond X-ray pulsars (AMXPs) at
low $\dot{M}$, this is essentially negligible (but is included in all
calculations for completeness).

\begin{figure}
 {\includegraphics[width=90mm]{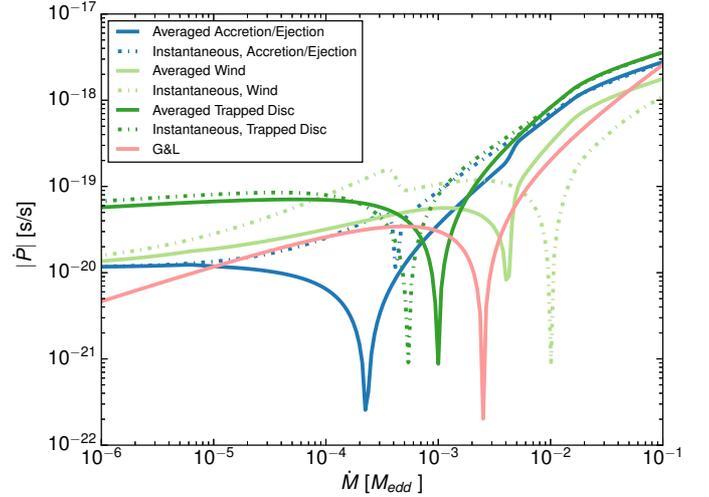}}
  \caption{Expected spin change (absolute value) as a function of the
    average accretion rate for an accreting millisecond pulsar with
    the canonical parameters given in table \ref{tab:refcoords} for
    different spin regulation models. The solid curves trace the net
    spin change after an outburst cycle (shown in Fig.~\ref{fig:mdott}
    and scaled to \Mavg). The dashed lines show the same spin change
    curves considering only the average accretion rate. (Note that
    $|\dot{P}|$ is plotted; there is a sign change at the singularity
    and at low $\dot{M}$ the star spins down.) \label{fig:jdot}}
\end{figure}

The solid lines in Fig.~\ref{fig:jdot} show the spin change induced in
a weak-field neutron star as a function of accretion rate for the
accretion/ejection model (dark blue), stellar wind (light green), and
trapped disc (dark green). The spin change predicted by GL is also
shown in pink for comparison. At high accretion rates all the
solutions converge, adding angular momentum at a rate
$\dot{M}(GMr_{\rm m})^{1/2}$. At low accretion rates, the effect of
the trapped disc becomes clear: unlike in the accretion/ejection
picture or when there is a stellar wind, the trapped disc is able to
keep spinning down the star efficiently.

\section{Method Summary}
During an accretion outburst, the spin period derivative is expected
to change with the accretion rate. Here I want to investigate whether
the net spin change across the whole accretion/quiescent cycle is the
same as predicted from the cycle-averaged accretion rate. The method
used to calculate the equilibrium spin period and timescale for
magnetic stars going through accretion outbursts is described in
detail below. In brief: I first define an accretion outburst profile,
$\dot{M}(t)$ with average accretion rate \Mavg, and calculate the
time-dependent torque $\dot{J}(\dot{M}(t))$ for a given spin-down
model model, which is then averaged over the outburst to get
$\langle\dot{J}(\dot{M}(t))\rangle$. In general, this can be very
different from $\dot{J}(\langle\dot{M}\rangle)$ i.e. the torque from
the time-averaged accretion rate. The solid and dot--dashed curves in
Fig.~\ref{fig:jdot} show how the net torque on a star changes with
accretion rate, either considering the effect of accretion bursts (the
solid lined `average' curves) or not (the dot--dashed 'instantaneous'
curves).

To calculate the spin evolution of a star, I then calculate a series
of $\langle\dot{J}(\dot{M}(t), P_*)\rangle$, i.e. the angular momentum
change as a function of accretion rate for a wide range of stellar
spin periods (as shown in Fig.~\ref{fig:jdot_ae}, where I have plotted
the $\dot P$, the stellar spin period change rather than the analogous
$\dot J$). I then use this series of curves to evolve the star's spin
for a given average accretion rate until it converges to a fixed spin
period.

The time it takes to converge is the spin equilibrium time, \teq\ and
the final spin period at convergence is the spin equilibrium period,
\peq. \teq\ and \peq\ for different torque models and outburst
properties can then be compared to analytic estimates without
accounting for accretion bursts (i.e. considering
$\dot{J}\langle\dot{M}\rangle$).

\subsection{Note on units, conversions for different types of systems}

Where possible, all results in this paper are given in terms of
scale-invariant variables that can be applied to different magnetized
accreting stars -- neutron stars with high ($\sim 10^{12}$G) magnetic
fields (X-ray pulsars) and low ($\sim 10^{8}$G) magnetic fields
(accreting millisecond X-ray pulsars or non pulsating neutron stars
with low-mass companions), magnetic white dwarfs, and TTauri
stars. Table~\ref{tab:refcoords} gives typical values of $B_*$, $P_*$,
$\dot{M}$, $r_{\rm m}$, etc. for different astronomical objects.
\begin{table*}
\caption{Adopted canonical values for different types of
  astrophysical systems.}
\label{tab:refcoords}
\begin{tabular}{lccccccccc}
\hline
Star & Mass & Radius & $B_*$ & $I_*$ & $\langle\dot{M}\rangle$ & $P_{\rm eq,0}$ & $T_{\rm eq,0}$ & $R_{\rm m}/R_*$ & $R_{\rm c}/R_*$ \\
  & ($M_\odot$) & (cm)  & (G) & (g cm$^{2}$) & ($M_\odot$ yr$^{-1}$)  & & (yr) & &  \\
\hline
Pulsar & 1.4 & $10^6$ & $10^{12}$ & $10^{45}$ & $1.4\times10^{-9}$  & 0.7s & $2\times10^5$ & 130 & 170\\
AMXP & 1.4 & $10^6$ & $10^{8}$ &  $10^{45}$ & $1.4\times10^{-11}$  & 0.002s & $6\times10^9$ & 2.5 & 2.7\\
Intermediate polar & 0.6 & $10^9$ & $10^{6}$ &  $10^{50}$ & $1.6\times10^{-10}$  & 1200s& $2\times10^6$ & 14 & 13\\
TTauri star & 0.5 & $1.4\times10^{11}$ & $2\times10^{3}$ & $4\times10^{54}$ & $5\times10^{-8}$  & 2 d & $3\times10^5$ & 2.7 & 3.4 \\
\end{tabular}
\end{table*}

Each torque model has several numerical parameters that I explore in
individual subsections. For the `canonical' versions of each model, I
adopt the following parameters:
\begin{itemize}
\item{in all models, $\xi = 0.4$ (the numerical factor modifying
  equation \ref{eq:rmGL} to set the location of \rin)}
\item{for the `accretion/ejection' and `wind' models, I adopt a smoothing length
    $\Delta r_2/r = 0.1$, while the wind has an assumed outflow rate
    of $0.1\dot{M}$}
\item{in the `trapped disc' model, $\Delta r/r = \Delta r_{2}/r =  0.1$}.
\end{itemize}
 The results in Section \ref{sec:results} use the stellar parameters
 of a millisecond X-ray pulsar (AMXP) listed in Table
 \ref{tab:refcoords}. In Sections~\ref{sec:AMXP}--\ref{sec:YSO} I
 discuss in more detail the simulation results specific different
 types of magnetic star and implications for spin evolution in these systems.
 
\subsection{Modelling the accretion outburst}
 I use a simple fast-rise/exponential-decay function to model an
 accretion outburst. This model has two free parameters: the duration
 of the outburst and the ratio between maximum and minimum accretion
 rate:
\begin{equation}
\label{eq:mdott}
  \dot{M}(t) = e^{\sqrt{2/F_{\rm t}}}e^{-1/10t - 10t/F_{\rm t}} + \dot{M}_{\rm min},
\end{equation}
where $\sim F_{\rm t}/5$ is the decay time and $\dot{M}_{\rm min}$ is the
quiescent accretion level through the disc. The ratio between
outburst maximum and quiescence is then:
\begin{equation}
  \frac{\dot{M}_{\rm max}}{\dot{M}_{\rm min}} \simeq
  \frac{e^{-0.6/\sqrt{F_{\rm t}}}}{\dot{M}_{\rm min}} + 1.
\end{equation}
\begin{figure}
  {\includegraphics[width=90mm]{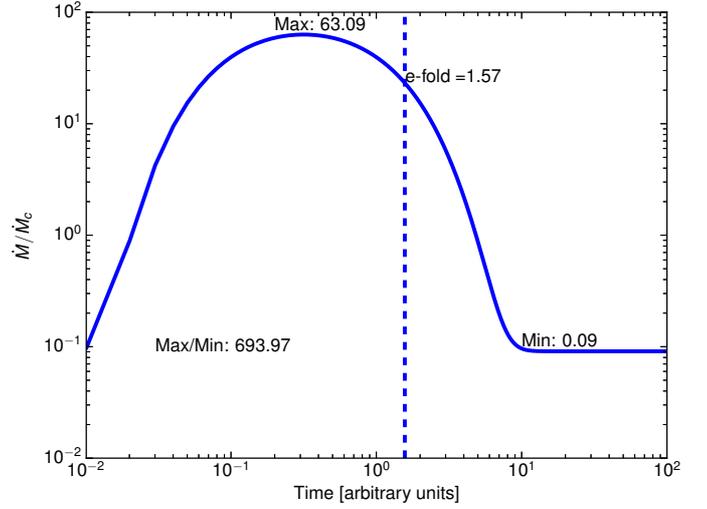}} \caption{Accretion
    rate as a function of time, scaled to $\dot{M}_{\rm c}$, the
    nominal equilibrium accretion rate for different spin
    periods. Accretion outbursts are modelled by a 'fast-rise, exponentially 
    decaying' function above a quiescent accretion level, where the
    outburst duration and ratio of peak $\dot{M}$ to quiescent
    $\dot{M}$ are free parameters.\label{fig:mdott} }
\end{figure}
The outburst duration is arbitrarily set to 100, so that $F_{\rm t}/5$
is a rough measure of the duration of the outburst (the rise times are
assumed to happen extremely rapidly for simplicity). The light curve
is then renormalized to 1 (which is why $\dot{M}_{\rm min}$ does not
always match the actual quiescent $\dot{M}$ in some figures).

The `canonical' burst profile adopted in this paper is shown in
Fig.~\ref{fig:mdott}. In this model $\dot{M}_{\rm min} =
0.0014~\dot{M}_{\rm max}$, $\dot{M}_{\rm min} = 0.1$, and the outburst
duration (defined as when the accretion rate is within $1/100e$ of
maximum) is $F_{\rm t} = 10$. In Section~\ref{sec:duration_amp} I
explore how changing the outburst duration and amplitude changes the
spin period. I have also explored other outburst shapes to confirm
that changing the functional form of the outburst (e.g. to a linear
rise and decay function) makes only small quantitative changes in the
results, provided there is a consistent outburst duration.

\subsection{Calculating $\dot{J}$ for different stellar spin periods and $\dot{M}$}

From the accretion profile $\dot{M}(t)$ and a given model for the
instantaneous angular momentum exchange between disc and star
(Sections \ref{sec:AccEj}--\ref{sec:trapped}), the net angular
momentum exchange over an entire outburst is calculated for different
average accretion rates. $\dot{J}(\dot{M})$ is a function of stellar
spin as well as the current accretion rate.  This can be made
scale-invariant by scaling the accretion rate by the `critical'
accretion rate:
\begin{equation}
  \label{eq:mdotc} 
\dot{M}_{\rm c}= \xi r_{\rm c}^{-7/2}\mu^2(GM_*)^{-1/2},
\end{equation}
i.e. the accretion rate at which $r_{\rm c} = r_{\rm m}$. Written this
way, $\dot{J}$ can be written as a function of a single variable,
$\dot{J}(\dot{M}/\dot{M}_{\rm c})$.

 Additional physical effects break the scale invariance of
 $\dot{J}(\dot{M}/\dot{M}_{\rm c})$. At very low $\dot{M}$, spin down
 can become dominated by magnetic dipole radiation for an AMXP. At
 high $\dot{M}$ neutron stars can reach the Eddington limit
 ($\dot{M}_{\rm Edd} \simeq 8.7\times10^{17}$~g~s$^{-1}$ for a
 $1.4M_{\odot}$ neutron star), which I assume is the maximum accretion
 rate on to the stellar surface (thus limiting spin up). Finally, in
 both TTauri stars and AMXPs at high $\dot{M}$ the accretion flow can
 crush the magnetosphere and fall directly on the star. In this case
 the torque on the star can be very different (see
 e.g. \citealt{1991ApJ...370..597P,1991ApJ...370..604P}). Since this
 is not the focus of this current paper, I simply assume that when the
 calculated \rin $< R_*$, the angular momentum added to the star at a
 rate of $\dot{M}(GM_*r_*)^{1/2}$. I similarly do not put a limit at
 the breakup frequency for the stars, since at very high $\dot{M}$
 where this is most relevant magnetospheric accretion will have ceased
and it is not clear how angular momentum is regulated.

 \begin{figure}
  {\includegraphics[width=90mm]{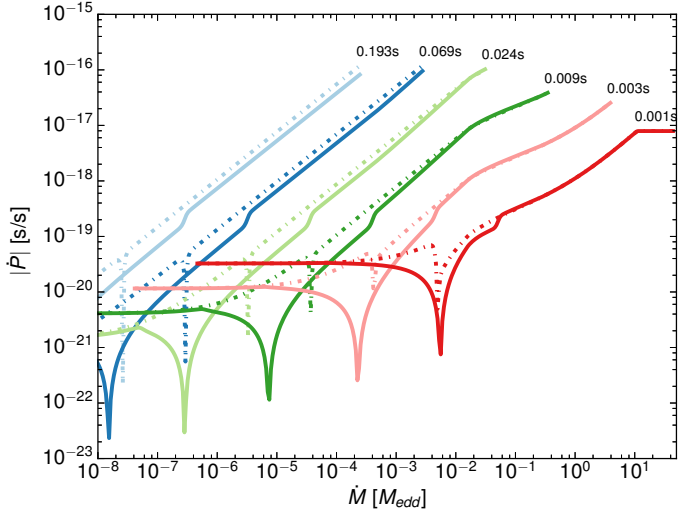}}
 \caption{\label{fig:jdot_ae} Spin change as a function of $\dot{M}$
   for the accretion/ejection scenario. The different curves represent
   different stellar spin periods (labelled above each curve). The
   dashed lines show the expected spin-down profiles for $\dot{P}$ as
   a function of a given instantaneous accretion rate. The minimum
   occurs at the spin equilibrium point, where the star switches from
   spin-up to spin-down. The solid curves instead plot
   $\langle\dot{P}(\dot{M}(t))\rangle$, the change in the net
   $\dot{P}$ as a function of $\dot{M}$ integrated over the entire
   outburst (the solid curves). Considering an outburst can
   dramatically alter the spin change of the star, decreasing
   $\dot{P}$ considerably over a wide range of $\dot{M}$ and shifting
   the `spin equilibrium' accretion rate systematically lower. The
   simple form of the function is broken by dipole spindown at low
   $\dot{M}$, and the Eddington limit at high $\dot{M}$.}
\end{figure}

Fig.~\ref{fig:jdot_ae} shows the spin rate change, $\dot{P}(\dot{M})$
of the `canonical' neutron star in response to the accretion/ejection
torque model. In all figures I plot $\dot{P}$ versus $\dot{M}$ rather
than $\dot{J}$ versus $\dot{M}$ to make the figures easier to relate to
observations. The two quantities are related by:
\begin{equation}
  \dot{P} = -\frac{\dot{J}P^2_*}{2\pi I_*},
\end{equation}
where $P_*$ is the stellar period and $I_*$ the star's moment of
inertia. Each curve shows $\dot{P}(\dot{M})$ for a different spin
period (ranging between $P_* = 0.001$--$0.2$s). The dot--dashed dashed
lines show the spin change as a function of the average accretion rate
($\dot{P}(\langle\dot{M}\rangle)$; equation~(\ref{eq:jdotAE})). The
solid lines show $\langle\dot{P}(\dot{M})\rangle$, the spin derivative
averaged over the entire accretion outburst and quiescence (calculated
using the outburst profile given in Fig.~\ref{fig:mdott}).

The singularity marks the `spin equilibrium' point, where the net
torque on the star is zero and there is no spin change. At accretion
rates less than equilibrium the star spins down, while for higher
$\dot{M}$ it spins up. The spin-down at very low $\dot{M}$ is
dominated by pulsar dipole radiation (relevant for AMXPs). At high
accretion rates, the spin-up torque on the star levels off as the
inner edge of the accretion disc first touches the star and then the
outburst accretion rate reaches the Eddington
limit\footnote{\cite{2005MNRAS.361.1153A} explicitly considered
  accretion from an Eddington-limited disc on to a neutron star, and
  found a different angular momentum exchange rate than the one
  presented here. This could be incorporated into this work, but is
  currently omitted for simplicity, and to make it easier to translate
  between different types of magnetic star.}.

The differences between the dashed and solid sets of curves are
clear. First  $|\dot{P}|$ over an outburst is much
smaller than the instantaneous spin change for a large range of
accretion rates close to equilibrium.  This means the total torque on
the star is considerably smaller than simple calculations would
predict, which significantly increases the time the star takes to
reach spin equilibrium (\teq). The second effect is to shift the
equilibrium accretion rate to a lower $\dot{M}$, which means that
\peq\ is significantly faster than equation~(\ref{eq:jdotAE}) would
predict for a given average accretion rate.

Fig.~\ref{fig:jdot} shows $\dot{P}(\langle\dot{M}\rangle)$ and
$\langle\dot{P}(\dot{M})\rangle$ for the different models of stellar
angular momentum loss presented in
Sections~\ref{sec:AccEj}-\ref{sec:trapped}, using the `canonical'
model parameters introduced above. Again, solid curves show the net
$\dot{P}$ averaged over an outburst, while the dashed curves show the
instantaneous $\dot{P}$ for a given accretion rate. For comparison,
the model of GL (as approximated by \citealt{2014MNRAS.437.3664H}) is
overplotted in dark green.

The main difference between the accretion/ejection model (blue), wind
(light green) and trapped disc model (dark green) is seen at low
$\dot{M}$. In a trapped disc at low accretion rates the torque from
the disc/field interaction remains strong, whereas for the other two
models spin-down is dominated by dipole radiation. The shape of the
accretion/ejection and wind models are asymmetric around the
equilibrium minimum. The minimum is also shifted relative to the
curves showing the instantaneous $\dot{P}(\dot{M})$, corresponding to
a different equilibrium spin period. In contrast, the trapped disc
shows a much more modest change in shape, although the equilibrium
$\dot{M}$ is also significantly shifted.  These differences underscore
the intrinsic model dependence in inferring properties of the star
(like the B field) from the assumption of `spin equilibrium' in a
magnetic star.

\subsection{Spinning the star towards equilibrium}
\label{sec:spin_equilibrium}
\begin{figure}
  {\includegraphics[width=90mm]{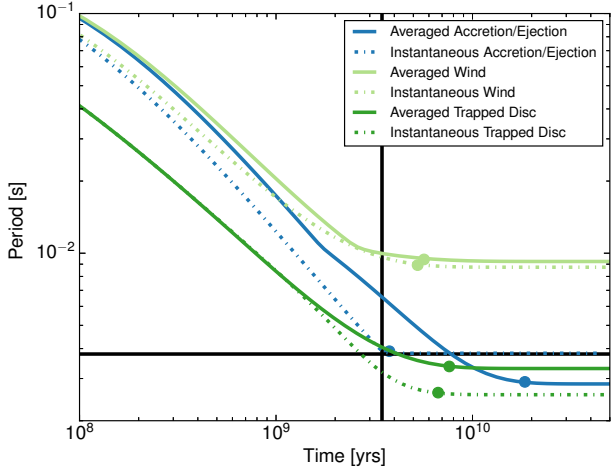}
    \caption{The final spin evolution of the `canonical' neutron star,
      for the same initial parameters but different torque models
      (accretion/ejection, wind and trapped disc). The horizontal
      solid line marks the expected equilibrium spin period, \peq\,
      while the vertical linemarks \teq. The circles indicated the
      point $(T_{\rm eq}, P_{\rm eq})$ for each
      model.\label{fig:spin_change}}}
\end{figure}

Finally, I use the set of curves $\langle\dot{J}(\dot{M},\dot{M}_{\rm
  c})\rangle$ calculated in the previous section to find \peq\ and
\teq\ for a given angular momentum model. For a given average
$\dot{M}$ and assumed stellar moment of inertia $I_*$, I evolve the
spin rate of the star in time in response to the torque
$\langle\dot{J}(\dot{M},\dot{M}_{\rm c})\rangle$ using a fifth-order
Runge--Kutta integration scheme implemented as `{\tt dopri5}' in the
{\tt scipy} library until the spin period converges.

I define the spin equilibrium time, \teq, as the evolution time for
the spin period from $(1\pm\epsilon)P_0$ to $(1\pm\epsilon)P_{\rm f}$,
where $\epsilon = 10^{-3}$ is an arbitrary parameter and $P_0$,
$P_{\rm f}$ are the initial and final spin periods). The $\pm$ sign is
used appropriately depending on whether the star spins up or down as a
result of accretion. The calculated spin equilibrium period is then
defined as the spin period at \teq.

Fig.~\ref{fig:spin_change} shows the final stages of spin evolution
for an AMXP accreting at $\dot{M}_0 = 2\times10^{-4} \dot{M}_{\rm Edd}$ . The
different colours correspond to the different torque models (as in
Fig.~\ref{fig:jdot}), for both $\langle\dot{P}(\dot{M})\rangle$
(solid) and $\dot{P}(\langle\dot{M}\rangle)$ (dot--dashed). The closed
circles indicate the numerically calculated \teq\ and \peq\ for each
spin-down curve.

Vertical and horizontal lines mark the analytic predictions for the
equilibrium spin period, $P_{\rm eq,0}$ calculated based on \Mavg:
\begin{eqnarray}
\label{eq:Peq}
P_{\rm eq,0} &\sim& \frac{2\pi}{\sqrt{GM_*}}r_{\rm m}^{3/2}\\
\nonumber P_{\rm eq,0} &\simeq& 3.2 {\rm ms} \left(\frac{\xi}{0.4}\right)^{3/2}\left(\frac{M_*}{1.4M_\odot}\right)^{-5/7}\\
\nonumber  &&\left(\frac{\mu}{10^{26} {\rm G~cm^3}}\right)^{6/7}\left(\frac{\dot{M}}{2\times10^{-4}\dot{M}_{\rm Edd}}\right)^{-3/7},
\end{eqnarray}
and the characteristic spin-down time, $T_{\rm eq,0} \simeq \dot{P}/P$:
\begin{eqnarray}
\label{eq:Teq}
T_{\rm eq,0} &\sim&\frac{2\pi I_*}{P_{\rm eq}\dot{M}(GM_*r_{\rm m})^{1/2}}\\
\nonumber T_{\rm eq,0} &\simeq& 3.5\times10^{9} {\rm yr}\left(\frac{I_*}{10^{45} {\rm g~cm^2}}\right)\left(\frac{\xi}{0.4}\right)^{-2}\left(\frac{M_*}{1.4M_\odot}\right)^{2/7}\\
\nonumber &&\left(\frac{\dot{M}}{2\times10^{-4}\dot{M}_{\rm Edd}}\right)^{-3/7}\left(\frac{\mu}{10^{26}{\rm G~ cm^3}}\right)^{-8/7}.
\end{eqnarray}

Both \peq\ and \teq\ can be substantially different from the values
given by equations~(\ref{eq:Peq}) and (\ref{eq:Teq}).  \peq\ and \teq\
are sensitive to both the torque model used and the properties of the
outbursts (their amplitude and duration), as well as the location of
the inner disc edge, and the specific details of the torque model
itself.

\begin{figure}
  {\includegraphics[width=90mm]{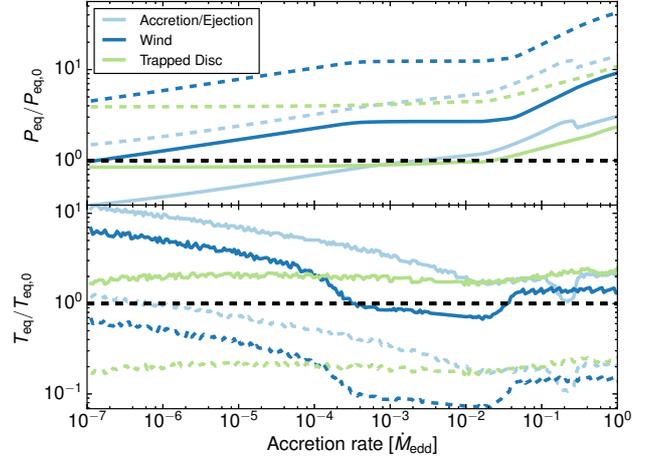}}
  \caption{\label{fig:SEP_SET} \peq\ (top) and \teq\ (bottom) for each
    torque model, accounting for outbursts, divided by the \peq\ and
    \teq\ predicted by simple analytic estimates (see text for
    details). The solid line compares the two quantities with the
    values obtained by taking $\langle\dot{M}_{\rm tot}\rangle$
    (i.e. the total averaged accretion rate), while the dashed line
    estimates \teq\ and \peq\ considering only average accretion rate
    during outburst \Mout.}
\end{figure}

Fig.~\ref{fig:SEP_SET} shows \peq\ and \teq\ (calculated and
illustrated in Fig.~\ref{fig:spin_change}) as a function of $\dot{M}$.
To emphasize the difference between simple estimates and more
realistic calculations, both \peq\ and \teq\ here scaled by
equations~(\ref{eq:Peq}) and (\ref{eq:Teq}). The solid curves scale
the results using the average $\dot{M}$, while the dashed curves show
the results scaled to \teqo\ and \peqo\ calculated using the {\em
  outburst} accretion rate \Mout, as is sometimes done in the
literature (e.g. \citealt{2014MNRAS.437.3863K})~{\footnote {In this
    case the predicted \teq\ will be increased by $1/f$, where $f$ is
    the fraction of time spent in outburst.}}.

Neither \Mavg\ nor \Mout\ gives a reliable measure of \teq\ and
\peq\ at all accretion rates, with deviations of up to an order of
magnitude in the estimated value.  \peq\ and \teq\ tend to follow a
power-law relationship with \Mavg\ for low accretion rates, which is
broken at higher accretion rates by $\dot{M}_{\rm Edd}$.  In
particular, the conclusion that a stronger magnetic field leads to a
slower star (equation~(\ref{eq:Peq}) is significantly complicated by
outbursts, and depends to some extent on the dominant angular momentum
loss mechanism.

Additionally, \teq\ is considerably longer than expected, which could
mean that spin equilibrium is unlikely.  This is especially true for
millisecond pulsars, where $T_{\rm eq}$ can easily stretch to $10^9$
yr at $\sim 10^{-3}\dot{M}_{\rm Edd}$, and TTauri stars, which have
outburst cycles that could be a significant fraction of \teqo\ (see
sec. \ref{sec:YSO}).

\section{Results}
\label{sec:results}

\subsection{Changing outburst duration and amplitude}
\label{sec:duration_amp}
Accreting stars show a wide range of outbursting behaviour, with
dramatic differences in outburst durations and amplitudes. Here 
investigate how this changes the spin evolution, first by varying the
outburst duration, then by varying its amplitude. In all cases the
average accretion rate is kept constant. 

\begin{figure}
  {\includegraphics[width=90mm]{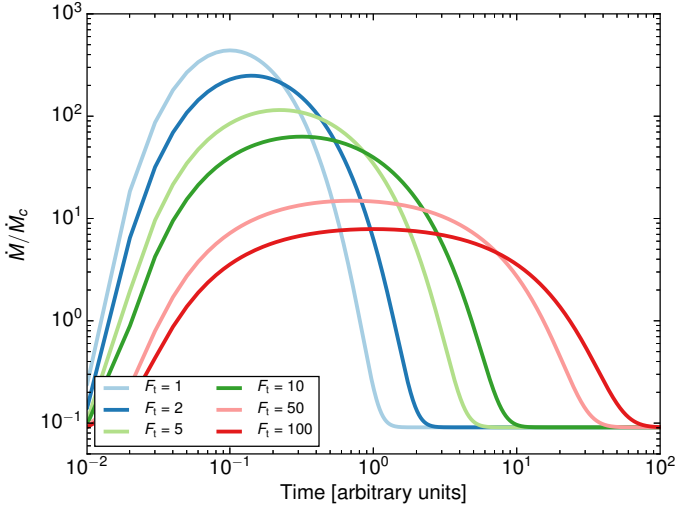}}
  \caption{\label{fig:MvsT_ft} Outburst accretion profiles for
    different outburst duration, keeping the net accretion rate fixed
    (so that shorter outbursts have larger maxima). The outburst
    duration varies from 0.2 to 20 per cent of the accretion cycle.}
\end{figure}

\begin{figure}
  {\includegraphics[width=90mm]{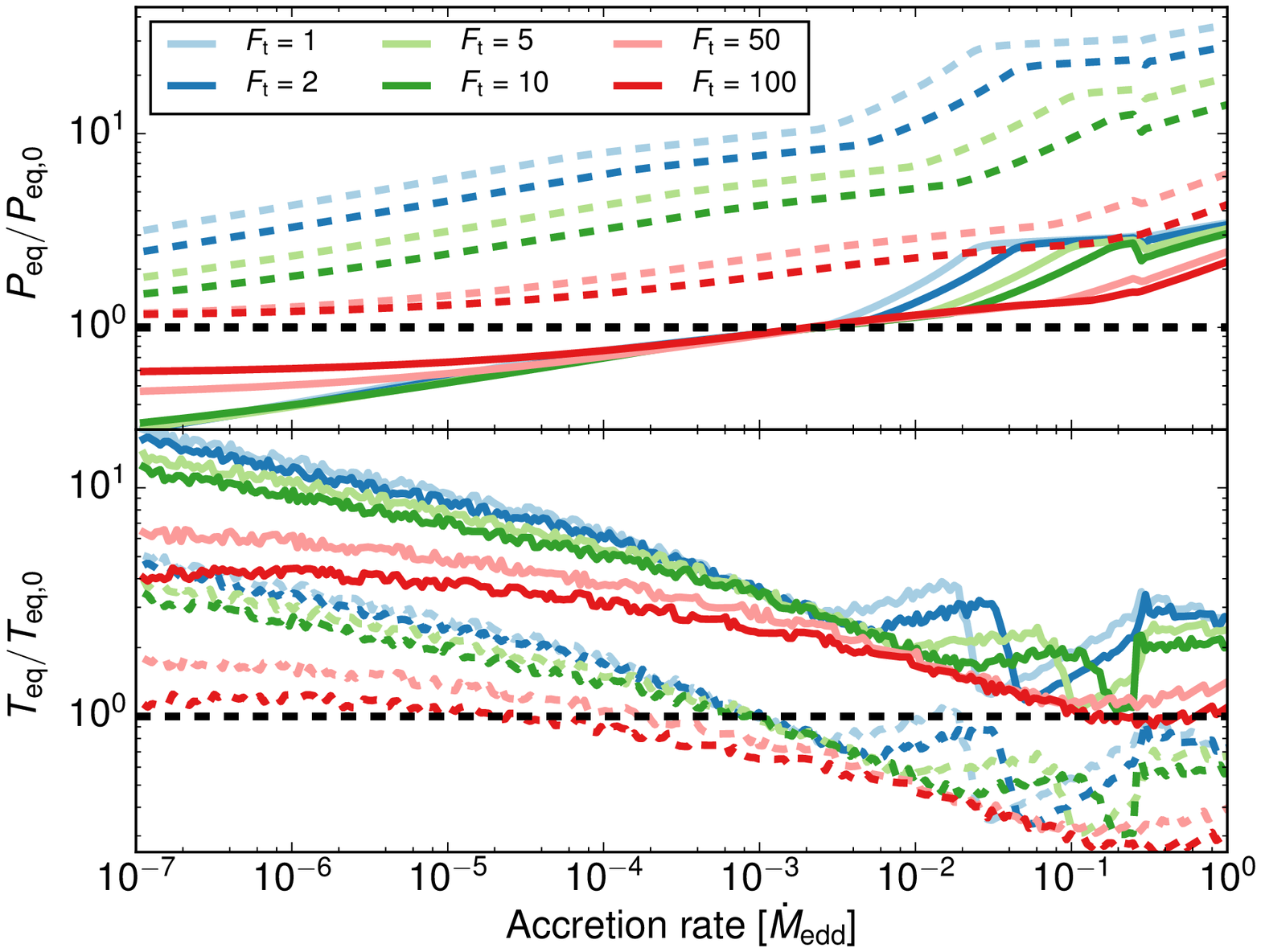}}
  \caption{\label{fig:SEP_SETft1} Effect of changing the outburst
    duration for the accretion/ejection model. For plot details see
    Fig.~\ref{fig:SEP_SET}. As can be seen from the figure, neither
    \Mout\ nor \Mavg\ can be used to give accurate estimates for
    \peq\ or \teq, particularly for short outbursts (small $F_{\rm
      t}$). The spin-equilibrium times are generically much longer
    than would be expected and the actual final spin periods are
    either considerably shorter (considering \Mavg) or longer (\Mout)
    than expected. The results are similar for the wind model.}
\end{figure}  
\begin{figure}
  {\includegraphics[width=90mm]{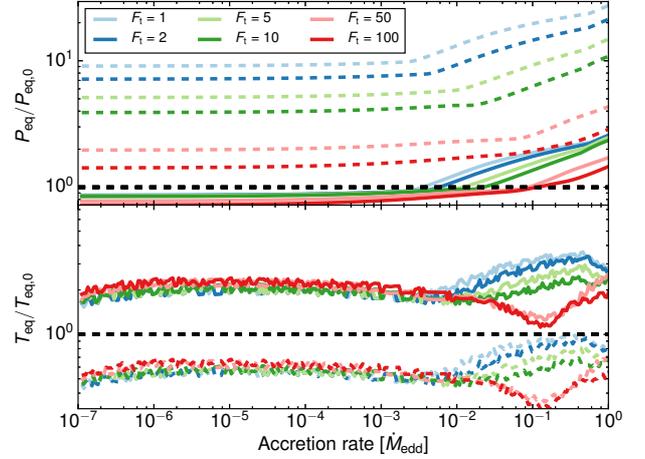}}
  \caption{\label{fig:SEP_SETft2} Same as Fig.~\ref{fig:SEP_SETft1}
    but for trapped disc model, in which spin down continues during
    quiescence. The \teq\ is best predicted from the outburst
    accretion rate, although the final spin period is mainly
    determined by \Mavg.}
\end{figure}

Fig.~\ref{fig:MvsT_ft} shows $\dot{M}(t)$ profiles for outbursts of
different duration, corresponding to $F_{\rm t} = [1,2,5,10,50,100]$
(cf. equation (\ref{eq:mdott})), with outburst durations
$T_{\rm out} = [0.5,0.9,2,4,15,26]$ and
$\dot{M}_{\rm max}/\dot{M}_{\rm min} =
[4800,2700,1300,700,160,90]$. The resulting \peq\ and \teq\ curves for
the accretion/ejection and trapped disc models are shown in
Figs~\ref{fig:SEP_SETft1} and \ref{fig:SEP_SETft2}. (The wind model
shows the same qualitative behaviour as the accretion/ejection model, except
that \peq\ is in general 2--3$\times$ longer, and \teq\ is typically
within 50\% of \teqo, as in Fig.~\ref{fig:SEP_SET}.)

The differences between the two models are clear. In the
accretion/ejection model, \teq\ is sensitive to the outburst duration
and neither \Mout\ nor \Mavg\ give a reliable analytic estimate of
\teq\ and \peq. Moreover, the difference between estimating \peq\ from
\Mout\ versus \Mavg\ is largest for very short outbursts.  For mean
accretion rates above $\sim10^{-4} \dot{M}_{\rm Edd}$, \Mavg\ gives a
fairly reliable estimate for \peq, while using \Mout\ predicts spin
periods \peq\ $\sim10\times$ shorter than the spin period from
considering outbursts. However, using \Mout\ (corrected for the time
spent in outburst) generally gives a more reliable estimate for
\teq\ than \Mavg.

Interestingly, \peq\ increases above \peq0\ for very high accretion
rates. Although this effect is modest, it only requires that accretion
is Eddington-limited, not that spin-up efficiency is reduced
\citep{2005MNRAS.361.1153A}. Both effects together may significantly
limit the maximum pulsar frequency without the need for other physical
processes such as gravitational wave emission
(e.g. \citealt{1998ApJ...501L..89B,2003Natur.424...42C,2012ApJ...746....9P}).

The trapped disc model in contrast is able to spin down the star at
very low $\dot{M}$ so \peq\ and \teq\ are fairly insensitive to
changes in \Mavg. Using \Mavg\ in equations (\ref{eq:Peq}) and
(\ref{eq:Teq}) thus gives the best estimate for \peq\ and \teq; \peq
$\sim0.8P_{\rm eq,0}$, while \teq\ is about twice as long. As
\Mavg\ increases the spin period becomes longer than predicted,
because the accretion rate (and hence spin up) becomes
Eddington-limited. \teq\ increases relative to \teqo, although again
by a modest factor. In summary, the trapped disc model matches well
with analytical predictions provided \Mavg\ is used rather than \Mout.

There is a wide variation in outburst durations in accreting
stars. Observed duty cycles for outbursting Be X-ray binaries are
typically 5--20 per cent
(\citealt{2011Ap&SS.332....1R,2014MNRAS.437.3863K}, see also
Section~\ref{sec:Be}) while for AMXPs the rate is more likely 2--3 per
cent per cent \citep{2015ApJ...805...87Y}, and for TTauri stars is
essentially unknown (\citealt{2015ApJ...808...68H} assume $\sim1$ per
cent).

\begin{figure}
  {\includegraphics[width=90mm]{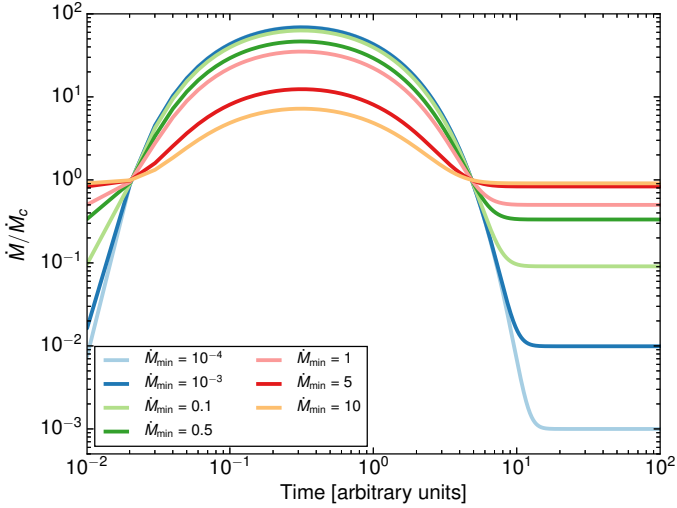}}
  \caption{\label{fig:MvsT_mmean} Outburst accretion profiles changing
    the mean quiescent level of accretion from $10^{-4}-10$, keeping
    the outburst duration the same. Since the accretion profile is
    normalized so that $\langle\dot{M}\rangle=1$, $\dot{M}_{\rm
      max}/\dot{M}_{\rm min} = [7\times10^4,7000,700,70,15,8]$ as the
    quiescent $\dot{M}$ increases.}
\end{figure}

The amplitude of the outbursts can also significantly affect the spin
evolution. Fig.~\ref{fig:MvsT_mmean} shows $\dot{M}(t)$ for a constant
outburst duration but amplitude variations over five orders of
magnitude, $\dot{M}_{\rm max}/\dot{M}_{\rm min} = [7, 15, 70, 150,
  700, 7000, 7\times10^4]$. Observed outburst amplitudes vary from
$\sim$10 to 1000 (Be X-ray binaries; \citealt{2011Ap&SS.332....1R}),
$\sim 10^3$ (TTauri stars, assuming they all undergo FU Ori-type
outbursts; \citealt{2015ApJ...808...68H}; Section~\ref{sec:Be}) and
$10^4-10^5$ (AMXPs).
 
\peq\ and \teq\ for the accretion/ejection model are shown in
Fig.~\ref{fig:SEP_SETmean,S}. Both of these values show somewhat
complicated behaviour at different $\dot{M}$ and for different
outburst amplitudes.  Predictably, for smaller outburst amplitudes
\peq\ and \teq\ stay very close to \peqo\ and \teqo. For larger
contrasts ($\dot{M}_{\rm max}/\dot{M}_{\rm min} > 100$), the
equilibrium spin periods are much shorter than would be predicted from
\Mavg, but are also generally significantly longer than predicted by
\Mout. Likewise, \teq\ is not well-predicted from either \Mavg\ or
\Mout. As the accretion rates become Eddington-limited (which happens
at progressively lower \Mavg\ for increasing outburst amplitude),
\peq\ and \teq\ in all cases start to more closely match
predictions. These results again emphasize the limitations of
inferring quantities like the stellar magnetic field from assumptions
of spin equilibrium and steady accretion.

Similar to changing the outburst duration, changing the outburst
amplitude has a minimal effect on \peq\ and \teq\ in the trapped disc
scenario (Fig.~\ref{fig:SEP_SETmean,T}). There is a modest increase in
\peq\ and \teq\ for larger outburst amplitudes, which is the same for
all \Mavg\ except at the highest accretion rates.

 \begin{figure}
    {\includegraphics[width=90mm]{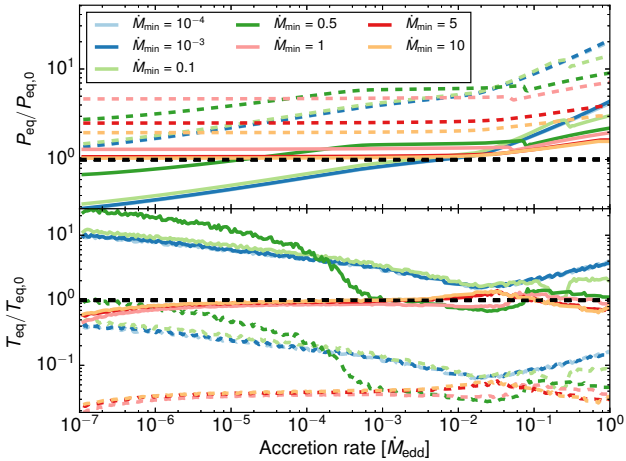}}
    \caption{\label{fig:SEP_SETmean,S} Same as Fig.~\ref{fig:SEP_SET},
      but now changing the outburst amplitude $\dot{M}$, using the
      accretion/ejection torque model. The different colours correspond to the
      outbursts with different total amplitudes
      (Fig.~\ref{fig:MvsT_mmean}).  As for the cases shown in
      Figs~\ref{fig:SEP_SETft1} and \ref{fig:SEP_SETft2}, the
      \teq\ is somewhat well-predicted by the outburst $\dot{M}$, while the
      \peq\ is better predicted from \Mavg.}
\end{figure}

\begin{figure}
    {\includegraphics[width=90mm]{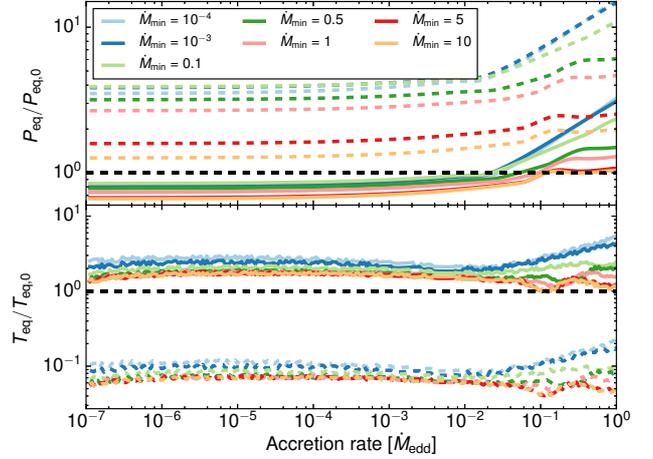}}
    \caption{\label{fig:SEP_SETmean,T} Same as Fig.~\ref{fig:SEP_SET},
      but now changing the outburst amplitude $\dot{M}$, using the
      trapped disc model. The different colours correspond to the
      outbursts with different total amplitudes
      (Fig.~\ref{fig:MvsT_mmean}). }
\end{figure}

\subsection{The effect of changing $r_{\rm m}$ ($\xi$) on spin evolution}
\begin{figure}
  {\includegraphics[width=90mm]{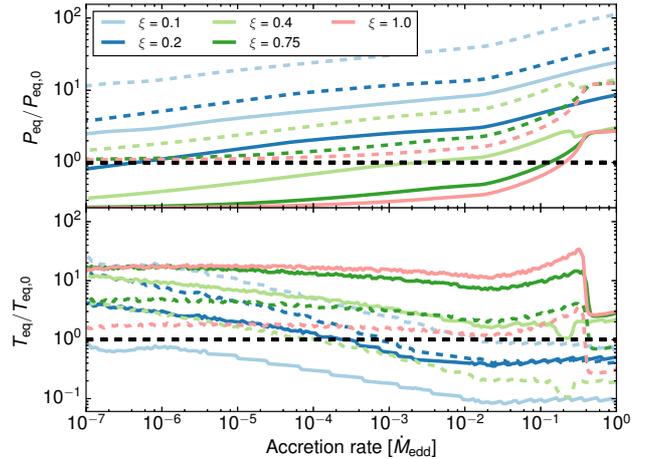}}
  \caption{Effects of changing inner radius of the disc ($\xi$) on the
    accretion/ejection model. The plot shows the \peq\ (top) and \teq\ (bottom)
    as a function of \Mavg\ for different $\xi$ values, analogous to
    Fig.~\ref{fig:SEP_SET}.\label{fig:SEP_SETxi1}}
\end{figure}
\begin{figure}
  {\includegraphics[width=90mm]{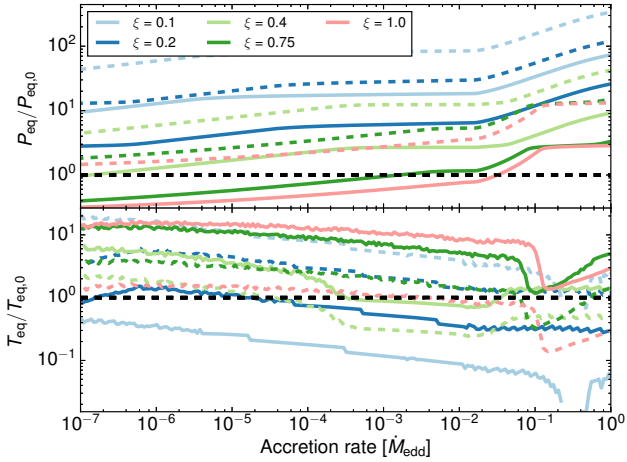}}
  \caption{Effects of changing inner radius of the disc ($\xi$) on the
    wind model. The plot shows the \peq\ (top) and \teq\ (bottom)
    as a function of \Mavg for different $\xi$ values, analogous to
    Fig.~\ref{fig:SEP_SET}.\label{fig:SEP_SETxi2}}
\end{figure}
\begin{figure}
  {\includegraphics[width=90mm]{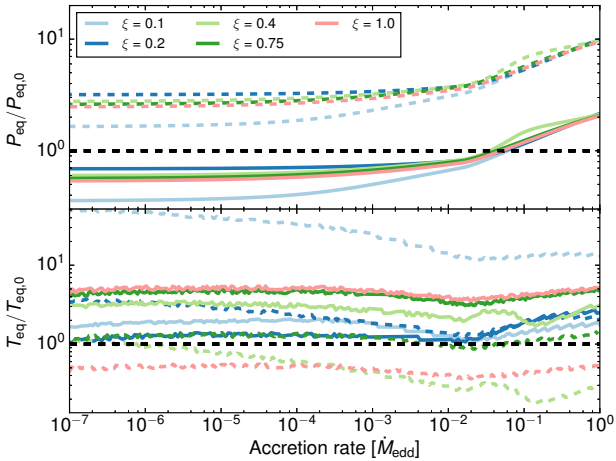}}
  \caption{Effects of changing inner radius of the disc ($\xi$) on the
    trapped disc model. The plot shows the \peq\ (top) and \teq\ (bottom)
    as a function of \Mavg for different $\xi$ values, analogous to
    Fig.~\ref{fig:SEP_SET}.\label{fig:SEP_SETxi3}}
\end{figure}

As discussed in Section~\ref{sec:rm}, the location of the inner edge
of the disrupted disc could depend on the structure of the accretion
flow, magnetic field configuration, and efficiency of coupling between
the disc and the field. The location of \rin\ also determines
$\dot{M}_{\rm c}$, the `critical' accretion rate that marks the
transition from predominantly spin-up to spin-down.  All these effects
can lead to changes in the spin-down efficiency. Here, the uncertainty
in \rin\ is parametrized by $\xi$ in equation~(\ref{eq:rmGL}), which
is usually assumed to lie between $\xi \sim 0.4$ and 1
(\citealt{2002apa..book.....F}).   In this paper I explore a
larger range for $\xi \sim 0.1$--1. This is motivated in part by the
fact that changing $\xi$ changes $\dot{M}_{\rm c}$, so that there can
be a considerable amount of spin down at high accretion rates. This is
motivated by the results of \cite{2013A&A...550A..99Z} and others, who
find significant outflows and angular momentum loss even at large
accretion rates, which means that the equilibrium accretion rate will
be significantly larger than expected. 

Although the location of \rin\ is only uncertain by a factor of a few,
the strong dependence of $\dot{M}$ on \rin\ means that the {\it
  accretion rate} for which $r_{\rm m} = r_{\rm c}$ can be uncertain
by a factor $\xi^{-7/2} = 300$ and $P_{\rm eq} \propto
\xi^{-3/2}\dot{M}^{-3/7}$ can lead to a 30-fold difference in spin
period.

The results of this section demonstrate how different $\xi$ affect the
long term spin evolution of an outbursting star for the three
different torque models, using the canonical outburst and stellar
parameters introduced in
Section~\ref{sec:mag_acc}. Figs~\ref{fig:SEP_SETxi1}--\ref{fig:SEP_SETxi3}
show \peq$(\dot{M})$ and \teq$(\dot{M})$ for the accretion/ejection,
wind, and trapped disc models, with results scaled by
equations~(\ref{eq:Peq}) and (\ref{eq:Teq}) setting $\xi = 0.4$.

Figs~\ref{fig:SEP_SETxi1} and \ref{fig:SEP_SETxi2} show the results
for the accretion/ejection and wind models. In both models
\peq\ increases roughly linearly for decreasing $\xi$, while \teq\ is
roughly proportional to $\xi$. The smaller $\xi$, the smaller
\rin\ for a given $\dot{M}$. Since $\dot{P} \propto \dot{M}r_{\rm
  m}^{1/2}$, for small $\xi$ less angular momentum is added, limiting
spin up. Even neglecting $\xi < 0.4$, \peq\ and \teq\ are uncertain by
$\sim 8\times$ with the largest deviations at high $\dot{M}$. The
uncertainty in $\xi$ also introduces uncertainty in whether using
\Mavg\ or \Mout\ will more accurately predict \peq, which again
underscores the difficulty in constraining physical parameters from
assumptions of spin equilibrium.

In the trapped disc picture (Fig.~\ref{fig:SEP_SETxi3}), \teq\ shows a
much weaker dependence on $\xi$ and \peq\ is nearly independent of
it. This is because \rin\ in a trapped disc always stays close to the
corotation radius and continues to extract angular momentum
efficiently for all $\xi$. In general, however, \teq\ is longer by
$\sim$ 2--5$\times$ than would be expected from analytic estimates,
again raising the question about whether spin equilibrium is a good
assumption, particularly for systems (like protostars or AMXPs) in
which the total duration of accretion could be comparable to \teq.
This is discussed further in Sections~\ref{sec:AMXP}--\ref{sec:YSO}.

\subsection{Exploring uncertainties in the `accretion/ejection' scenario}
\label{sec:transition}
\begin{figure}
  {\includegraphics[width=90mm]{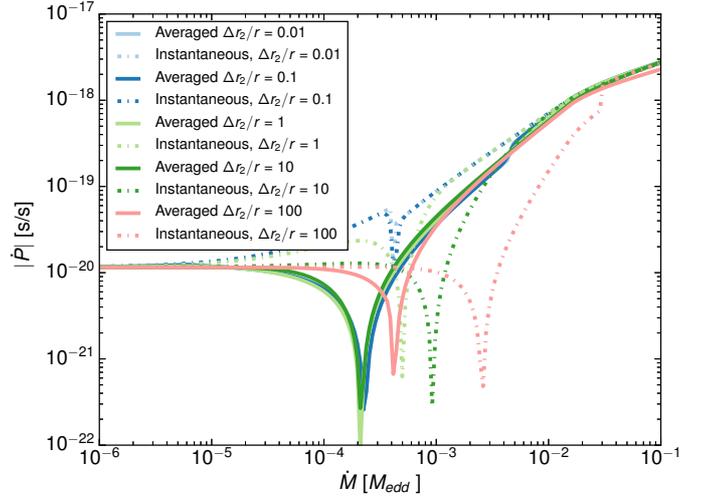}}
  \caption{\label{fig:jm4smoothGL} The spin down rate as a function of
    accretion rate in the accretion/ejection model, for a star with
    $P_* = 0.003$s. The different dashed curves show the effect of
    changing the smoothing parameter $\Delta r_2/r$ in
    equation~(\ref{eq:jdotAE}). This changes the range of $\dot{M}$ in
    which the inner disc is close to $r_{\rm c}$ and there is
    simultaneously spin up and spin down, before moving to a `true'
    ejection/accretion state. The solid lines show the torque averaged
    over an outburst, demonstrating that except for the very largest
    values of $\Delta r_2/r$, the net torque is barely affected by the
    width of the transition region between propeller and accretion.}
\end{figure}

\begin{figure}
  {\includegraphics[width=90mm]{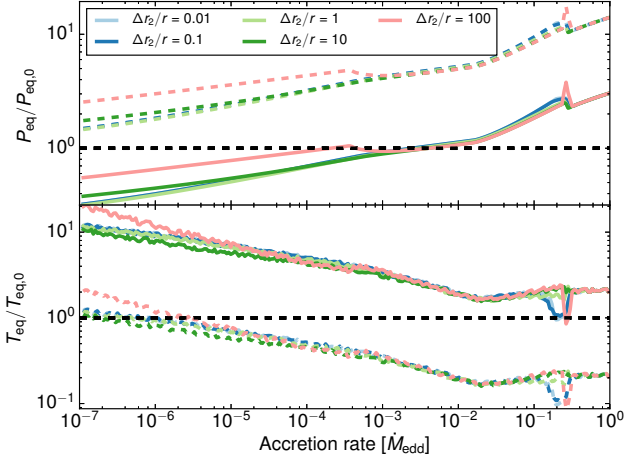}}
  \caption{\label{fig:sep_set4smoothGL} \peq\ and \teq\ curves
    corresponding to the different curves in
    Fig.~\ref{fig:jm4smoothGL}. As is clear from the figure, the
    transition width between the propeller and accretion regimes makes
    very little difference in determining the final spin period or
    spin equilibrium time.}
\end{figure}

The simplest `accretion/ejection' picture for magnetospheric accretion
is one in which gas is accreted at high $\dot{M}$, expelled in a
centrifugally-launched `propeller' outflow at low $\dot{M}$, and shows
both infall and outflow for a range of intermediate $\dot{M}$.  This
model is approximated by equation~(\ref{eq:jdotAE}).  Numerical MHD
simulations generically indicate that the disc--field interaction is
time-dependent, and show some accretion and outflow for all
$\dot{M}$. As a result there can be significant angular momentum
loss while the star is actively accreting (as observed by
\citealt{2013A&A...550A..99Z}), and some residual accretion
reaches the star even in the strong `propeller' regime, but how much
or how the accretion/outflow efficiency with \Mavg\ is
unclear. \cite{2015MNRAS.449.2803D} used numerical simulation results
to quantify the ejection efficiency (fraction of gas expelled in an
outflow) as a function of accretion rate, which suggest propeller
efficiencies of up to $\sim95$ per cent in the strongest propeller
simulations. In equation~(\ref{eq:jdotAE}) the transition from
`propeller' to `accretion' is parametrized by $\Delta r_2$, which
gives the range of \rin\ which have both accretion and ejection (or
equivalently, the range of \Mavg\ where this is the case, $\Delta
\dot{M} \equiv -7/2 \Delta r_2/r \dot{M}$).
  
Fig.~\ref{fig:jm4smoothGL} shows how the star's spin rate changes with
$\dot{M}$ for different values of $\Delta r_2$ in the `accretion/ejection'
torque model. To emphasize how little $\Delta r_2$ affects $\dot{P}$,
the figure shows $\dot{P}(\dot{M})$ for $\Delta r_2$ across four
orders of magnitude: $\Delta r_2/r = 10^{-2..2}$. As in previous
figures, the solid lines show the outburst-averaged $\dot{P}$,
$\langle\dot{P}(\dot{M})\rangle$, while the dot--dashed lines show the
`instantaneous' spin rate change, $\dot{P}(\langle\dot{M}\rangle)$.

As the figure shows, $\Delta r_2$ has a very strong effect on the
instantaneous torque on the star, generally suppressing both spin-up
and spin-down efficiency and increasing $\dot{M}_{\rm c}$ (the
equilibrium accretion rate) as $\Delta r_2$ increases. However, the
effect of increasing $\Delta r_2$ on the {\em outburst averaged} (and
therefore long term) torque is minimal. As a result, \peq\ and
\teq\ are essentially unaffected by changes in $\Delta r_2$, as shown
in Fig.~\ref{fig:sep_set4smoothGL} because the accretion rate in
outburst declines so rapidly that spin down is only efficient
(e.g. $\sim3\times10^{-4}$ for $\Delta r_2/r = 0.01$) for a short
time.

This result indicates that even though the `accretion/ejection' model adopted
here is quite simplified, the detailed form of $\dot{J}(\dot{M})$ does
not strongly influence the long term evolution of an outbursting
magnetic star. 

\subsection{Changing the wind amplitude}
In general, the conclusions of the previous section also apply to the
accretion wind model (and explains why the results of this model are
generally similar to the accretion/ejection one). In both cases, spin-down is
inefficient when outbursts are considered because the accretion rate
drops too quickly for the star to spend much time in the spinning-down
phase. 

\begin{figure}
  {\includegraphics[width=90mm]{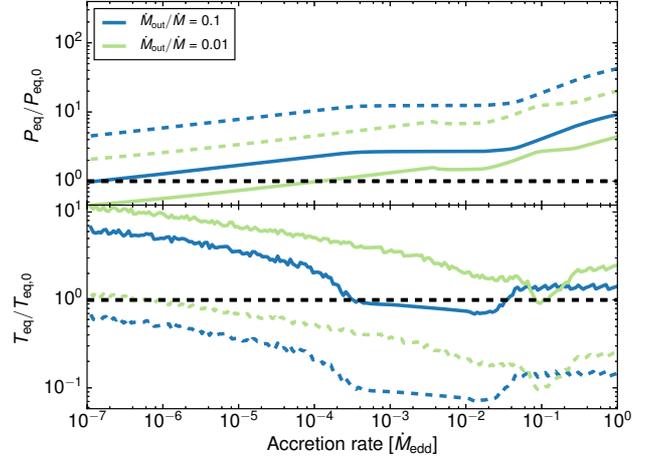}}
  \caption{\label{fig:Wind_strength} Same as Fig.~\ref{fig:SEP_SET}
    for two different outflow rates from a stellar wind. When the
    outflow rate is high ($\sim 0.1\dot{M}$) \peq\ becomes
    significantly slower than predicted, by a factor of $\sim 2-10$,
    depending on the accretion rate. A more significant effect is to
    decrease \teq\ to $\sim$\teqo\ across a wide range of accretion
    rate.}
\end{figure}

Fig. \ref{fig:Wind_strength} shows \peq\ and \teq\ for a strong
($\dot{M}_{\rm out} = 0.1 \dot{M}$) and weak ($\dot{M}_{\rm out} = 0.1
\dot{M}$) wind. For small outflow rates ($\sim 0.01\dot{M}$), the
qualitative behaviour is essentially the same as for the
accretion/ejection model, which indicates a stellar wind must be very
strong (and requires a significant amount of accretion energy for
launching) in order to significantly affect the spin evolution of the
star. As long as $\dot{M}_{\rm out}$ remains a free model parameter,
it is difficult to say whether a stellar wind model is really a viable
source for angular momentum loss. However, high-field neutron stars
(which have the largest magnetospheres of any magnetically accreting
star) show no indications of strong outflows, either as a radio source
or through interaction with their environments, which might be
expected for $\langle\dot{M}_{\rm out}\rangle \sim 0.01-0.1
\dot{M}_{\rm Edd}$.

\subsection{Exploring the trapped disc model}
\begin{figure}
  {\includegraphics[width=90mm]{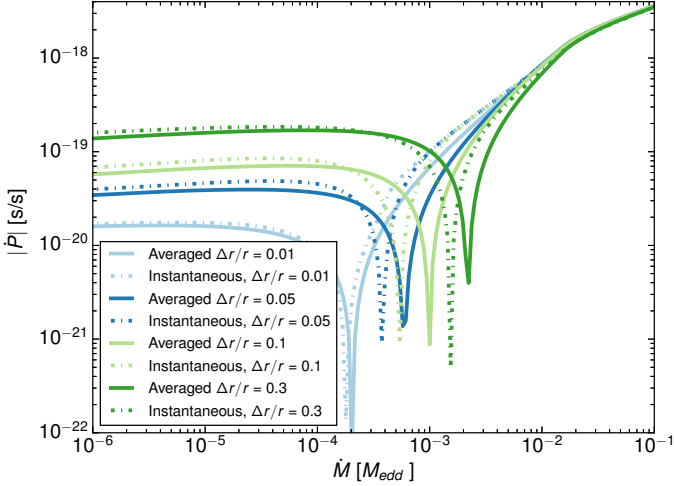}}
    \caption{\label{fig:trapped_dr} Spin-down rate for the canonical
      stellar parameters for the trapped disc model, where different
      coupling strengths are assumed (corresponding to the width of the
      coupled disc--field region, $\Delta r/r$). Unlike the
      accretion/ejection picture, spin regulation continues in
      quiescence, so that there is only modest difference between
      using the actual outburst profile and the time-averaged one.}
\end{figure}

\begin{figure}
  {\includegraphics[width=90mm]{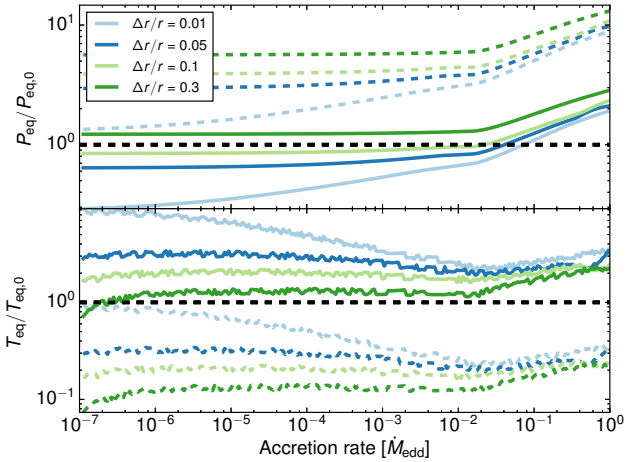}}
    \caption{\label{fig:setsep_dr} The ratio of \peq\ and \teq\ compared
      with the analytic prediction for a trapped disc with different
      coupling strengths (as shown in Fig.~\ref{fig:trapped_dr}).The
      \teq\ is not significantly affected by the change of coupling
      strength, but the equilibrium period increases for increasing
      coupling strength (since spin down is more efficient).}
\end{figure}

The trapped disc model also has two numerical parameters introduced in
DS10, which mainly reflect our ignorance of the details of the
disc--field interaction. The first is $\Delta r/r$, the width of the
coupled region between the disc and the magnetic field (which sets the
spin down efficiency of the interaction). The second is $\Delta
r_2/r$, which gives the range of \rin\ over which there is both spin
up and spin down (the same as in Section~\ref{sec:transition}). 

Figs~\ref{fig:trapped_dr} and \ref{fig:setsep_dr} show how changing
$\Delta r$ (i.e. the strength of the disc--field coupling) affects
$\dot{P}$ and the overall spin evolution of a
star. Fig.~\ref{fig:trapped_dr} shows $\dot{P}(\dot{M})$ curves for
the canonical trapped disc parameters with increasing coupling
strengths, $\Delta r = [0.01,0.05,0.1,0.3]$. Increasing $\Delta r$ by
$10\times$ increases $\dot{M}_{\rm eq}$ by $\sim5$, since a higher
$\Delta r$ increases the spin-down efficiency, so that \peq\ is slower
for the same \Mavg. There is a difference up to a factor 2 between the
instantaneous and averaged $\dot{M}_{\rm eq}$. For the canonical
outburst profile, the disc is generally {\em more} efficient at
spinning down the star than would be estimated from the na\"ive
formula balancing spin up and spin down.

Fig.~\ref{fig:setsep_dr} shows \peq\ and \teq\ as a function of
\Mavg\ for the torque models plotted in
Fig.~\ref{fig:trapped_dr}. Since the star can only spin down via
interactions with the disc, if $\Delta r$ is very small \teq\ will be
longer and \peq\ much shorter than expected (with increasing accuracy
as \Mavg\ increases). This effect is strongest for low accretion
rates, where the spin-down torques dominate. As the mean accretion
rate increases (and the magnetosphere becomes less important since the
disc reaches the star), the differences between strong and weak
spin-down decrease considerably. Finally, the figure shows (as is seen
throughout this paper), that for a trapped disc \Mavg\ is a better
predictor for \teq\ and \peq\ than \Mout.

\begin{figure}
  {\includegraphics[width=90mm]{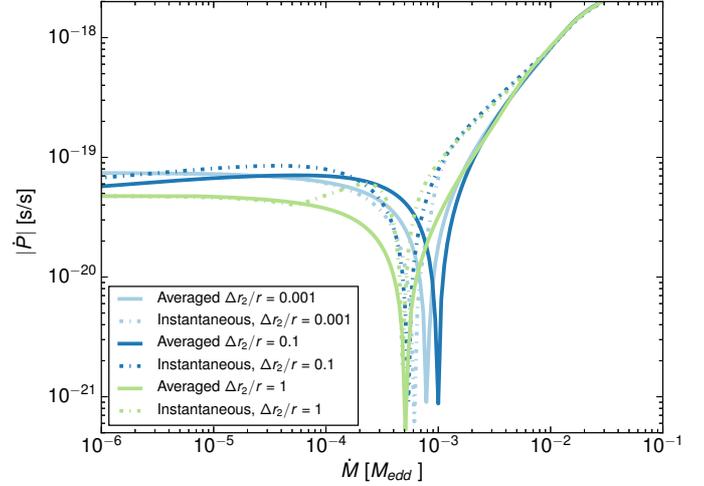}}
    \caption{\label{fig:trapped_sm} Spin-down rate for the canonical
      stellar parameters for the trapped disc model, for different
      values of softening length $\Delta r_2/r$ (analogous to
      Figs~\ref{fig:MvsT_mmean}-\ref{fig:sep_set4smoothGL}). As in
      the accretion/ejection scenario, changing the softening length
      has a modest effect on the net torque on the star.}
\end{figure}

Fig.~\ref{fig:trapped_sm} shows that increasing $\Delta r_2$ broadens
the region around $\dot{M}_{\rm eq}$ where the instantaneous torque is
reduced. The effect on the outburst-averaged $\dot{P}$ is somewhat
more subtle. For $\Delta r_2 = 0.1r$, $\dot{M}_{\rm eq}$ is {\em
  higher} than for both $\Delta r_2 = r$ and $\Delta r_2 = 0.01r$,
most likely related to the effects of accretion at large
$\dot{M}$. This effect is modest (a factor of about two in $\dot{M}$)
and might considerably change with the assumed outburst profile, so I
do not explore it further.

\begin{figure}
  {\includegraphics[width=90mm]{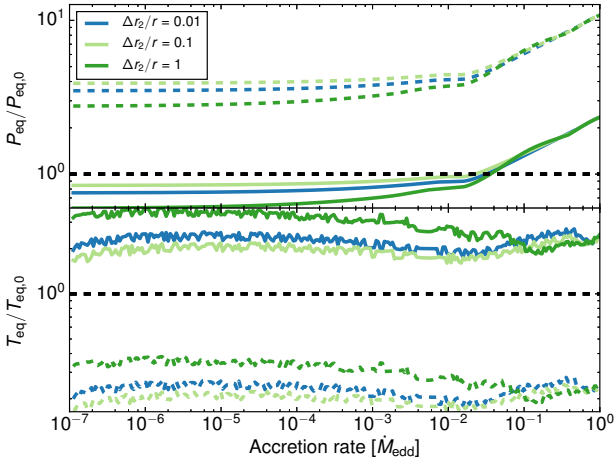}}
    \caption{\label{fig:setsep_sm} \peq\ and \teq\ for the system in
      Fig.~\ref{fig:trapped_sm}. Changing the softening length has
      only a modest effect on the spin period and little influence on
      the \teq.}
\end{figure}

The effects of changing $\Delta r_2$ on \peq\ and \teq\ are seen in
Fig.~\ref{fig:setsep_sm}. Here again the differences between models
are very modest and are close to the `expected' values; \teq\ varies
by $\sim2\times$ and \peq~$\lesssim~1.5~P_{\rm eq,0}$.

The conclusion of this and the previous section is that, compared with
the differences between torque models and the uncertainties in
accretion outburst details, the uncertainties in the individual torque
models have a modest effect on the spin equilibrium period or
spin-down/up timescale for the star. As long as there is efficient
spin down at some point, the different values of $\Delta r_2$ (total
magnitude of the torque when $P_*$ is close to \peq) does not
matter. The most important difference remains what happens during
quiescence -- whether there is substantial spin down (as in the
trapped disc model) or not (the accretion/ejection and wind models).

\section{Discussion}
This paper compares the long term spin evolution of magnetized stars
using different models for angular momentum regulation, explicitly
considering the effects of time-variable accretion. Here I briefly
discuss the consequences of the results presented in
Section~\ref{sec:results} for three (very different) types
of magnetically-accreting, outbursting stars: TTauri stars, AMXPs and
Be/X-ray binaries. I focus on two questions in particular:

\begin{enumerate}
\item{Can observations be used to distinguish between the trapped disc
  pictures and other models for spin regulation?}
\item{How does considering a variable accretion rate alter predictions
  of the observable properties of strongly magnetized accreting
  stars?}
 \end{enumerate}

\subsection{Accreting millisecond X-ray pulsars and LMXBs}
\label{sec:AMXP}
Accreting neutron stars with low magnetic fields ($\lesssim 10^8$ G)
with low-mass companions (low-mass X-ray binaries, or LMXBs) are
thought to be the progenitors of radio millisecond pulsars
(e.g. \citealt{1982Natur.300..728A}), spun up to millisecond spin
periods via accretion over hundreds of millions of years. AMXPs are a
subset of this group that show coherent pulsations and accretion
outbursts, with peak luminosities reaching $\sim 20$ per cent $L_{\rm
  Edd}$ (although most remain much fainter).

A second, partially overlapping subset of LMXBs have spin periods
measured through quasi-periodic oscillations (`burst oscillations'),
which are produced by localized, accretion-induced nuclear burning on
the star's surface modulated by the star's rotation. These are an
additional useful sample since the accretion rates in burst
oscillation sources can be significantly larger than AMXPs. The spin
periods inferred from burst oscillations are shorter on average than
in than AMXPs (although the sample size remains small;
\citealt{2014A&A...566A..64P}).  No periodicity has been detected in
all remaining LMXBs, despite some very deep searches
\citep{2015ApJ...806..261M}, which could mean that the magnetic field
in these sources is not strong enough to channel the accretion flow,
at least during the brightest phases of the outburst.

What limits the spin frequency of the millisecond pulsars? Despite
having relatively weak fields ($\sim 10^{8}$G; inferred from dipole
spin-down) and long accretion times (the donor star lifetime is often
$>1$ Gyr), the fastest radio millisecond pulsar has a spin period of
1.4ms \citep{2006Sci...311.1901H}, much longer than the theoretical
mass-shedding limit of $\sim 0.7$ms. Two possible mechanisms have been
proposed -- gravitational wave emission from a spin-induced quadrupole
moment or $r$-modes \citep{1998ApJ...501L..89B, 1999ApJ...516..307A}, or
the spin-down effects from the magnetic field/disc interactions
(e.g. \citealt{2012ApJ...746....9P}), but it has proven difficult to
definitively distinguish between them.

The spin distribution of radio millisecond pulsars peaks at a
significantly longer spin period than that of
AMXPs. \cite{2012Sci...335..561T} has recently suggested that the
difference in spin between the two populations could be significantly
affected by the evolution of the mass-transferring
companion. In this picture, AMXPs undergo a strong spin down during
the `Roche lobe decoupling phase' as companion stops filling its Roche
lobe so that the mass transfer rate to the pulsar
decreases. \cite{2012Sci...335..561T} estimated roughly 50 per cent of
the pulsar's angular momentum can be lost during this phase, during
which the average accretion rate drops by $\sim 3$ orders of
magnitude.

The present work challenges the assertion that a decrease in $\dot{M}$
will efficiently spin the star down particularly if the
accretion/ejection torque picture is the most relevant one. The
results of Section~\ref{sec:results} demonstrate the uncertainty in
estimating \peq\ and \teq: between uncertainties in the star-disc
interactions (e.g. the parameter $\xi$), the angular momentum loss
mechanism, and the presence of accretion outbursts, \peq\ and
\teq\ can both easily be uncertain by 10$\times$, even when the
physical parameters of the system ($\dot{M}$, $P_*$, $B_*$, $\dot{P}$)
are well constrained. \teq\ lengthens with declining $\dot{M}$, so
that as $\dot{M}$ decreases it takes progressively longer for the star
to reach a new spin equilibrium. The results in this paper show that
once outbursts are considered, \teq\ increases up to $10\times$.

Observations of AMXPs suggest that {\em none} of them are in spin
equilibrium with their time-averaged accretion rates (considering both
quiescence and outburst). On the other hand, systems with well
constrained spin derivatives show much less spin up during outburst
than might be expected from their luminosity
\cite{2012arXiv1206.2727P}, which could indicate spin
equilibrium. \cite{2008MNRAS.389..839W} finds the average
luminosity (including quiescence) in AMXPs varies between
$6\times10^{-5}$ and $0.02L_{\rm Edd}$ which implies (assuming radiative
efficiency and the average NS parameters adopted in this paper)
\peq\ $\sim 1$--$11$ms. AMXPs have observed spin periods between
$1.7-5.5$ms, with no obvious trend as a function of mean luminosity
(although some luminosities are uncertain by up to $10\times$ from
distance and bolometric uncertainties). In particular, the recently
discovered `transitional pulsars'
\citep{2009Sci...324.1411A, 2013Natur.501..517P, 2014MNRAS.441.1825B},
which switch between states of active accretion and radio pulsations,
all have relatively fast spin periods (1.7-3.9ms), despite extremely
low accretion rates during outbursts for two of the three systems. For
one of these sources, PSR J1023+0038, recent analysis of the X-ray
pulsations has found spin down during outburst is moderately larger
than dipolar spindown \cite{2016ApJ...830..122J} measured when the
accretion disc is absent in the radio-loud phase.

Nonetheless, RMSPs are observed to spin (on average) significantly
slower than AMXPs, which would be possible even if a trapped disc
remains present to spin down the star even at very low $\dot{M}$. The
spin-down in this case could happen gradually over the entire
long term decay phase of $\dot{M}$, rather than mainly being focused
at early times in the `Roche-lobe decoupling phase', as suggested by
\cite{2012Sci...335..561T}.

If the final large decline in $\dot{M}$ is not able to significantly
spin down most pulsars in their late accretion phase, the question of
what sets their maximum spin rate again becomes more urgent. In this
paper, the `canonical' \peq\ for an AMXP is about 0.4ms at
$\dot{M}_{\rm Edd}$, but all simulations with outbursting accretion
show slower rotation rates, typically by $\sim1.5$--$3\times$ but up
to $10\times$ in some cases. On the other hand, \teq\ at
$\dot{M}_{\rm Edd}$ is around 50Myr (and increases when outbursts are
considered). This is much shorter than the lifetimes of these systems,
and (based on the observed sample of LMXBs) is unlikely to dominate
the lifetime accretion rate of the star. As long as the star has a
$\sim10^{8}$G field, a lifetime average
$\dot{M} \sim 0.1-0.01\dot{M}_{\rm Edd}$ can limit the final spin
period to within observed values without invoking an additional
spin-down source like gravitational waves. (This is before considering
modifications to the spin-up rate, e.g. \citealt{2005MNRAS.361.1153A},
which may limit angular momentum transfer at high $\dot{M}$).

\subsection{Be/X-ray Binaries}
\label{sec:Be}

In strongly magnetized accreting neutron stars ($B\sim10^{12}$G),
dipole radiation is unimportant for spin regulation compared with spin
change from accretion, and the spin rate of the star is determined by
the interaction between the magnetic field and the accretion flow. The
observed spin distribution ($P_*\sim 1$--$1000$s) of these systems is
much larger than in AMXPs, and many systems are observed to spin up or
down considerably. However, many accreting high-field neutron stars
have high mass ($M>3M_\odot$) companions and are believed to mainly
accrete from a wind rather than a disc
(e.g. \citealt{1997ApJS..113..367B}), which is thought to give a much
larger spread in $P_*$ and $\dot{P}_*$ than results from disc
accretion.

A possible exception to this are Be/X-ray binaries, in which the
neutron star undergoes accretion outbursts when it passes through the
decretion disc of a companion Be star. Based on angular momentum
conservation arguments, \cite{2014MNRAS.437.3863K} argue that as the
pulsar passes through the Be star's disc most of the gas entering the
pulsar's sphere of influence will have too much angular momentum to
fall on to the star directly, implying that an accretion disc should
form around the neutron star.

The \textit{XMM--Newton} survey of the Small Magellanic Cloud (SMC) has
tracked pulsars in Be X-ray binaries in the SMC over the past 14
yr, providing a unique data set to test spin evolution models
\citep{2010ASPC..422..224C}.  \cite{2014MNRAS.437.3664H} and
\cite{2014MNRAS.437.3863K} argue that the small observed spin period
derivatives suggest spin equilibrium (or else extremely low magnetic
fields), and, if spin equilibrium is assumed, a surprisingly large
fraction of Be X-ray binaries in the SMC should have magnetar-strength
magnetic fields ($\sim 10^{14}$G). This is in contrast to systems in
our own Galaxy with similar spin rates and luminosities, which have
magnetic field estimates from cyclotron resonance emission lines on
the order $B\sim 10^{12}$G.

The conclusions of this paper suggest a somewhat different
interpretation of the observations discussed by
\cite{2014MNRAS.437.3863K}, which reduce (although do not completely
eliminate) the need for a very large magnetic field in most
pulsars. Be X-ray binaries are generally transient, so that their
average luminosity is much lower (typically several orders of
magnitude) than their luminosity in outburst. To estimate the magnetic
field, \cite{2014MNRAS.437.3863K} assume that the star is in spin
equilibrium {\em with the outburst accretion rate} (see
equation~\ref{eq:Peq}). This can be reasonable assumption if the
accretion/ejection model applies, since in quiescence the torque on
the star is strongly reduced. However, if a trapped disc remains
present during quiescence, the star continues to spin down, and it is
more accurate to consider the {\em average} $\dot{M}$ rather than the
outburst $\dot{M}$. (In fact, \textit{Fermi} observations of some
Be-X-ray binary systems indeed show that they spin down between
outbursts, see e.g. \citealt{2015arXiv150204461S}.)

To see how the results of this paper could affect estimates of $B_*$
in these systems, I calculate \peq\ (equation (\ref{eq:Peq})) assuming that
spin equilibrium has been reached, using \Mavg\ rather than \Mout\ (as
was assumed by \citealt{2014MNRAS.437.3863K}).  A rough estimate of
\Mavg\ for the stars in \citep{2014MNRAS.437.3863K} is given by:
\begin{equation}
\langle\dot{M}\rangle \simeq \dot{M}_{\rm out}F_{\rm out},
\end{equation}
where $\dot{M}_{\rm out} \simeq 0.01$--$0.2\dot{M}_{\rm Edd}$ (the
inferred accretion rate from the outburst luminosity), and $F_{\rm
  out} \simeq N_{\rm det}/N_{\rm obs}$ is the fraction of time spent
in outburst (the ratio between the number of detections to
observations). \cite{2014MNRAS.437.3863K} report 1-2 weekly
observations (I use 84 observations/yr) over a timespan ranging from
0.15 to 14 years, which corresponds to $F_{\rm out} \simeq 0.004$--$1$
($\langle F_{\rm out}\rangle \sim 0.06$) and $\langle\dot{M}\rangle
\simeq 7.5\times10^{-5}$-- $0.2~\dot{M}_{\rm Edd}$. This assumes that
the quiescent luminosity of these sources is at least 100$\times$
lower than in outburst, which seems roughly consistent with
observations (Coe, private communication).

\begin{figure}
  {\includegraphics[width=90mm]{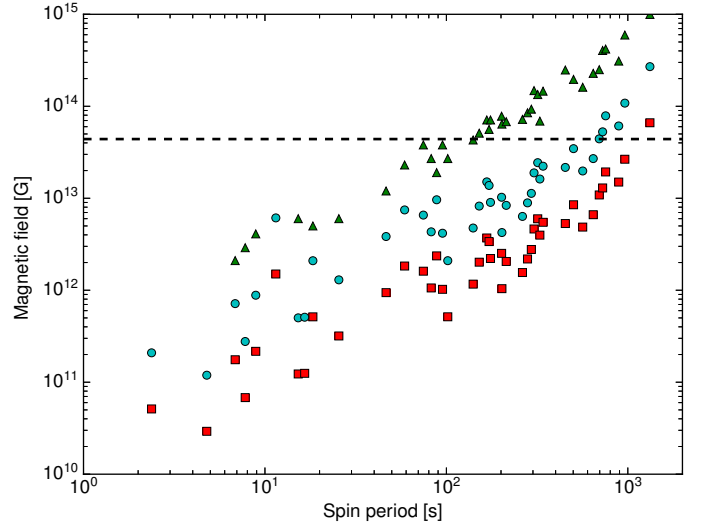}}
  \caption{Estimated magnetic field as a function of the spin period
    of Be X-ray binaries in the the SMC. The green triangles show the
    values calculated by \protect\cite{2014MNRAS.437.3863K} (which are
    roughly equivalent to using the reported outburst accretion rate
    and measured spin periods in equation~(\ref{eq:Peq})). The cyan
    circles show the same data set using the accretion/ejection model
    and the time-averaged accretion rate, while the red squares show
    the same results for the trapped disc model. The black dashed line
    shows the quantum critical field, $B_{\rm crit} =
    4.4\times10^{13}$G, where the cyclotron energy is comparable to
    the electron rest-mass energy, which is commonly used to define a
    `magnetar'. Using the time-averaged accretion rate to estimate $B$
    and assuming a trapped disc persists in quiescence obviates the
    need for magnetar-strength magnetic fields. \label{fig:BvsP}}
\end{figure}

Using equation~(\ref{eq:Peq}), the estimated \Mavg, and the reported
period for each pulsar from \cite{2014MNRAS.437.3863K}, I estimate a
revised magnetic strength, using either the accretion/ejection or
trapped disc model. For simplicity I choose the `canonical'
accretion/ejection and trapped disc models from
Section~\ref{sec:spin_equilibrium}, scaled to Be X-ray binary
parameters. The resulting $P_{\rm eq}$ is 0.9$P_{\rm
  eq,0}$\footnote{i.e., $P_{\rm eq}$ from equation~(\ref{eq:Peq})} for
a trapped disc, and $\sim 0.3P_{\rm eq,0}$ for the accretion/ejection
model. The resulting estimated magnetic fields are shown in
Fig.~\ref{fig:BvsP}. As is clear from the figure, using a
time-averaged accretion rate rather than the outburst one gives
systematically lower estimates for $B$ regardless of the torque model,
but if the systems are able to efficiently spin down during quiescence
(by transferring angular momentum into a disc), there is no need for
the majority of systems to harbour magnetar-strength fields. Since the
time-scales for reaching spin equilibrium in Be X-ray binaries are much
shorter than for either TTauri stars or XMSPs, this result provides
the strongest evidence for trapped discs around strongly magnetic
stars.

\subsection{Young Stellar Objects}
\label{sec:YSO}
TTauri stars also show strong evidence for spin regulation from
interaction with an accretion disc \citep{2007prpl.conf..479B}, and
most TTauri stars with discs spin well below their breakup rate,
despite the fact that they contract as they evolve. The different
mechanisms for angular momentum regulation discussed in this paper are
thus relevant for these stars as well. TTauri stars are also often
variable, showing variability on different time-scales. If the
variability is caused by large accretion rate variations on to the
stellar surface, then this should also affect the spin equilibrium
rate of the star, as discussed throughout this paper. TTauri stars are
more similar to AMXPs than high-field neutron stars, with a much
smaller magnetosphere that is probably completely crushed at high
$\dot{M}$.

Since variability time-scales are much longer in TTauri stars than
neutron stars, it is not straightforward to determine whether all
TTauri stars are variable. Recent work looking at variability has
found that the most common variability -- fluctuations on short
time-scales (days to weeks) is most likely due to variations on the
stellar surface that become apparent as the star rotates
\citep{2014MNRAS.440.3444C}. However, larger scale variability (which
is observed in a subset of TTauri stars) {\em is} attributed to
accretion rate fluctuations.

Variations of $\sim 10$--100 with time-scales of a few years are seen in
a subclass of TTauri stars known as `EXors', after the prototype, EX
Lupi \citep{2007AJ....133.2679H}. Even more dramatically, FU Ori-type
stars undergo luminosity increases of $\sim10^{3}$ times, and can
persist for 50--100+ years \citep{1996ARA&A..34..207H}. This paper is
particularly relevant for these last two subtypes, since very large
accretion rates should correspond to faster equilibrium spin
rates. There is growing evidence that EXors are a distinct class (or
alternately, evolutionary phase) of TTauri stars, so this phase may
not generally last long enough to be relevant for long term spin
rates. In contrast, the long quiescent time-scales conjectured for FU
Ori stars ($10^3$--$10^4$~yrs) mean that most or all TTauri stars
could pass through an extended FU Ori phase, which should then be
reflected in the final spin rate.

Comparing the estimated \teq\ for TTauri stars (see table
\ref{tab:refcoords}) with the predicted FU Ori outburst cycles shows
another important distinction between TTauri stars and magnetic
accreting compact objects: the duration of an outburst cycle is a much
larger fraction (up to 10\%) of the nominal equilibrium timescale
(which as discussed could be much longer). As a result assuming spin
equilibrium may not be valid.

Are the results of this paper consistent with observations of the spin
rates of TTauri stars? Assuming that most stars go through enough FU
Ori outbursts to reach spin equilibrium, the answer is sensitive to
how the spin rate of the star is regulated. As seen in Section
\ref{sec:results}, when a simple `accretion/ejection' picture is
assumed, the star tends to spin up to close to its outburst spin rate,
rather than the long term averaged one. For FU Ori stars, assuming a
duty cycle of between 0.1 and 1 per cent, \Mout $\sim
10^{-4}~\rm{M_{\odot}~yr^{-1}}$ versus
\Mavg\ $\sim10^{-7}$--$10^{-6}~\rm{M_{\odot}~yr^{-1}}$. For a typical
TTauri star, the accretion rate during outburst will be high enough to
completely crush the magnetosphere, so that the disc accretes through
a boundary layer directly on to the star. In standard accretion theory,
the star should then spin up to close to its breakup frequency
(although see discussion below). The high outburst accretion rate will
also presumably inhibit a magnetically driven wind from the stellar
surface, which will limit how efficiently a wind can regulate the
star's spin, and likely not be able to prevent the star from spinning
up. Na\"ively, one would then expect that TTauri stars in the FU Ori
outburst stage should be spinning significantly faster than
\peq\ estimated from observations, which is most likely $\dot{M}$ in
`quiescence'. This does not immediately seem to be the case, although
there may still be enough uncertainty in $B_*$ and the torque models
that distinguishing between the two scenarios could be difficult.

In contrast, a trapped disc spins down the star in the quiescent
state, and over time will bring the star into spin equilibrium with
its long term accretion rate. For a duty cycle of about 1 per cent, the
accretion rate is still fairly high ($10^{-6}\rm{M_\odot~yr^{-1}}$)
and corresponds to a faster spin than is observed (0.5--1 d). If the
duty cycle is shorter, the mean accretion rate can be close to the
quiescent one ($10^{-7}\rm{M_\odot~yr^{-1}}$), corresponding to a spin
period of a few days, which is roughly consistent with observed spin
periods.

These conclusions are also challenged by observational evidence that
suggests the magnetosphere \citep{2014MNRAS.437.3202J} and inner disc
of young stars \citep{2007prpl.conf..507N} are located well within
\rc. If these radius measurements are accurate it is somewhat
surprising even within the `standard' steady-state accretion model,
since it would suggest these stars are likely spinning up rapidly. It
may indeed suggest enhanced spin down torque at relatively high
accretion rates \citep{2013A&A...550A..99Z}. If most TTauri stars are
FU Ors in quiescence, the problem is even larger: one would expect
that the FU Ori events spin up the star even more, requiring even
stronger spin down at lower accretion rates.

There are several other possibilities for reconciling the high FU Ori
accretion rates with relatively long spin periods. One is that the FU
Ori phase of repeated outbursts may only occur for a subset of TTauri
stars, or that this accretion phase does not last long enough to bring
the star into spin equilibrium. This question can only be resolved
observationally. A second possibility is that accretion through a
boundary layer does not easily spin the star up to breakup. This has
been suggested in boundary layer calculations by
\cite{1996ApJ...467..749P} and more recently by new numerical and
analytical work \citep{2013ApJ...770...67B}. In the latter papers, the
authors find that angular momentum and energy in the boundary layers
are mainly transported via acoustic waves rather than an `anomalous
viscosity' as is typically assumed for both accretion discs and
boundary layers. \cite{2013ApJ...770...67B} instead find angular
momentum transport via waves can result in some outward transport
(i.e. back into the disc), as well as into the deep layers of the
star. Both these effects can limit how efficiently the star will spin
up, although by how much is not yet quantified.

However, without an additional very efficient and rapid source of
angular momentum loss, the results here studying spin change in
outbursts (both the expected final spin periods and the spin evolution
time), combined with results suggesting most discs are truncated well
within \rc, suggests that FU Ori phenomena are more likely a rare or
brief evolutionary state, and most observed TTauri stars are not in
the quiescent state of an FU Ori phase.

Finally, the conclusions from this section are somewhat preliminary,
since the models of spin evolution adopted in this paper do not
consider the radial contraction of the protostar during its lifetime,
which will make the star spin faster and hence require even more
angular momentum loss. While this is straightforward to include, it is
outside the scope of the current paper.

\subsection{Conclusions}
The results of this paper suggest that the long term spin evolution of
magnetic stars can be significantly affected by large-scale changes in
the mass accretion rate. In general, I find that by considering
accretion outbursts, stars take significantly longer to reach their
`equilibrium' spin period and that this spin period in general can be
significantly different (generally shorter, but not always) than would
be predicted from simple analytic arguments. The \peq\ and \teq\ are
sensitive to the disc--field interactions, the outburst duration, and
the transport mechanism that removes angular momentum from the star.

In particular, the commonly envisioned scenario, in which gas either
accretes on to the star or is expelled through a centrifugally launched
wind, requires that the average accretion rate stay fairly steady in
order to keep the star in near its predicted \peq. This is because the
spin down mechanism is only efficient at relatively high $\dot{M}$
(when the inner disc remains close to \rc). Interestingly, this
conclusion holds even for the more recent variants of this model, in
which there is both accretion and ejection across a large range of
$\dot{M}$ (section \ref{sec:transition}). Such a steady $\dot{M}$ is
inconsistent with the most widely accepted `ionization instability'
model for accretion outbursts, in which the accretion rate through the
disc varies by several orders of magnitude between outburst and
quiescence \citep{2001NewAR..45..449L}.

If a stellar wind (launched from the stellar surface but driven in
part by accretion power) can be launched, spin-down can remain
efficient as long as the mass outflow rate is high enough ($\sim$10
per cent $\dot{M}$). There is some evidence supporting this idea for
TTauri stars (e.g. \cite{2008ApJ...681..391M} and other works by those
authors), but the idea remains somewhat schematic and controversial
\citep{2011ApJ...727L..22Z}, and the outflow rate from the star itself
is difficult to constrain observationally. The `trapped disc' model
also has significant uncertainties, in particular the details of the
coupling between the disc and the star, and the width of the coupled
region (which sets the spin down efficiency), but has the distinction
of being able to spin down the star very efficiently even at low
$\dot{M}$. This could be very important in understanding the slow spin
rates of Be X-ray binaries and possibly the long term spin rates of
millisecond pulsars.

In AMXPs, the large difference between outburst and quiescence means
that accretion continues even when the cycle-averaged accretion rate
is in the `propeller' regime. This affects the conclusions of
\cite{2012Sci...335..561T}, in particular, the assertion that AMXPs
can efficiently spin down via a propeller during a `Roche lobe
decoupling phase' (where the mean accretion rate drops
rapidly). Observations indicate that AMXPs in general are {\em not} in
spin equilibrium with \Mavg. This could support the conclusion that
AMXPs are not the progenitor systems for the entire class of radio
millisecond pulsars \citep{2012arXiv1206.2727P} and therefore that their faster
average spin periods do not indicate a general spin evolution from one
population to the other; alternately it could suggest that a trapped
disc remains around the star even as \Mavg\ drops and continues to
spin the star down. Recent observations of transitional millisecond
pulsar systems \citep{2016ApJ...830..122J}, however, suggest that the
net spin down from an accretion flow is comparable to that from dipole
radiation.

In Be/X-ray binaries, considering the effects of outbursts changes the
estimates of magnetic field (calculated assuming spin equilibrium)
significantly. This conclusion applies even if an accretion/ejection
model is considered, but it is especially true if a trapped disc
remains present during quiescence. Considering these effects, the
estimated magnetic field strengths for Be/X-ray binary systems
(considering the large sample from the SMC,
\citealt{2010ASPC..422..224C}) is significantly lower than estimated
by \cite{2014MNRAS.437.3863K}, and in particular does not require in
magnetar-strength magnetic fields except for the slowest spinning
stars.

It is not currently clear to what extent all protostars undergo
repeated, large-scale outbursts, although at least a subset show
large-scale variability. In these systems the magnetosphere is likely
crushed by accretion during the outburst, so that the star should
accrete via a boundary layer. The outcome of this scenario is not
completely clear, but na\"ively one would expect that the final spin
rate of the star would be dominated by what happens during outbursts
\citep{1996ApJ...467..749P}. Observations of the innermost regions of
TTauri stars suggest that the inner disc and closed magnetosphere are
generally well within \rc, indicating that these stars are more likely
spinning {\em up} than spinning down after a large FU Ori-level
outburst. This fact, and the fact that observed spin rates are
generally much slower than breakup could then imply that either the
star accretes without spinning up efficiently during outburst, or that
FU Ori-type outbursts are not a universal or long-lasting phase of
star formation.  The results of this paper are preliminary though,
since they do not include the contraction (and necessary spin up) of
the star as it evolves, nor spin regulation via boundary layer
accretion.

{\bf Note:} After this paper appeared on the arXiv, a similar work,
focusing on transient accretion in AMXPs and considering only an
`accretion/ejection' model was also published
\citep{2017ApJ...835....4B}. The authors conclude based on their
analysis that gravitational waves may be required to prevent MSPs from
spinning to submillisecond periods during outburst. They broadly reach
the same conclusion as for the `accretion/ejection' case considered
here, namely, that stars should spin faster than predicted by the
average accretion rate because a propeller outflow is generally
inefficient, but do not find the same limit on spin period at the
highest accretion rate (from limited spin-up efficiency because the
source reaches $\dot{M}_{\rm Edd}$). Further investigation into what
happens at high accretion rates is ongoing, and this will include a
more detailed comparison with the results of that paper.

\section*{Acknowledgments}
Part of the inspiration from this work came from discussions at the
ISSI meeting: `The disk-magnetosphere interaction around transitional
millisecond pulsars', and was carried out at the Aspen Center for
Physics during the workshop `Universal Accretion: The Physics of Mass
Accretion on All Scales and in Diverse Environments' . This work was
funded by the NWO VIDI grant (PI:Patruno). I am particularly grateful
to Dr. Wynn Ho, Dr. Alessandro Patruno and Dr. Rudy Wijnands for their
careful reading of this manuscript and suggestions, as well as
interesting discussions with Malcolm Coe and Thomas Tauris.




\bibliographystyle{mnras}
\bibliography{magbib_2016} 

\bsp	
\label{lastpage}
\end{document}